\documentclass[preprint,12pt]{elsarticle}

\usepackage{amssymb}
\usepackage{amsmath}
\usepackage{xcolor}
\usepackage[colorlinks]{hyperref} 

\usepackage{float}
   \usepackage[hang,tight,footnotesize]{subfigure}
   \usepackage[displaymath,mathlines,left]{lineno}
   
   \usepackage{listings}
   \lstdefinestyle{myFortranStyle}{
      language=Fortran,
      stepnumber=1,
      numbersep=10pt,
      tabsize=6,
      showspaces=false,
      showstringspaces=false
   }
   \lstdefinestyle{myTextStyle}{
      language={},
      basicstyle = \footnotesize\ttfamily,
      numbers=left,
      frame = leftline,
      framesep = 0pt,
      framexleftmargin = 3pt,
      framerule = 3pt,
      rulecolor = \color{red!25!green!50!blue!75},
      numberstyle=\scriptsize\color{red!25!green!50!blue!75},
      numbersep = 10pt,
      tabsize = 3,
      xleftmargin=0.5in 
   }

   \usepackage{algorithm2e}
   \SetKwComment{Comment}{/* }{ */}
   \RestyleAlgo{ruled}
   \SetKw{KwBy}{by}
   \SetKwInOut{Input}{Input}
   \SetKwInOut{Output}{Output}
   \SetKw{Local}{Local variables:}
   
   \SetCommentSty{mycommfont}
   \usepackage[normalem]{ulem}
   \newcommand{\bI}{\mathbf{I}}
   \newcommand{\bU}{\mathbf{U}}
   
   \usepackage{multicol}
\newcounter{bla}

\journal{Computer Physics Communications}

\makeatletter
   \def\ps@pprintTitle{%
   \let\@oddhead\@empty
   \let\@evenhead\@empty
   \def\@oddfoot{}%
   \let\@evenfoot\@oddfoot}
\makeatother

\begin{document}

%% Viet added:
%   \linenumbers
%%

\begin{frontmatter}



\title{Exact nuclear pairing solution for large-scale configurations: I. The EP (v1.0) program at zero temperature}

\author[a,b]{Tran Quoc Viet \corref{author}}
\author[a,b]{Le Tan Phuc \corref{author}}
\author[a,b]{Tran Vu Dong}
\author[c,d]{Nguyen Ngoc Anh}
\author[a,b]{Nguyen Quang Hung}

\cortext[author] {Corresponding author.\\\textit{E-mail address:} tranquocviet5@duytan.edu.vn (T. Quoc Viet), letanphuc2@duytan.edu.vn (L. Tan Phuc).}
\address[a]{Institute of Fundamental and Applied Sciences, Duy Tan University, Ho Chi Minh City 70000, Vietnam}
\address[b]{Faculty of Natural Sciences, Duy Tan University, Da Nang City 55000, Vietnam}
\address[c]{Phenikaa Institute for Advanced Study (PIAS), Phenikaa University, Hanoi 12116, Vietnam}
\address[d]{Faculty of Fundamental Science, PHENIKAA University, Yen Nghia, Ha Dong, Hanoi 12116, Vietnam}

\begin{abstract}
%\begin{linenumbers} 

In this work, we present the ``EP code" (version 1.0), a user-friendly and robust computational tool. It computes the exact pairing eigenvalues and eigenvectors directly from the general nuclear pairing Hamiltonian, represented using SU(2) quasi-spin algebra with basis vectors in binary representation, at zero temperature for both odd and even deformed nucleon systems.
In this initial release, the sparsity and symmetry of the pairing matrix are exploited for the first time to quickly construct the pairing matrix. The ARPACK and LAPACK packages are employed for the diagonalization of large- and small-scale sparse matrices, respectively. In addition, the calculation speed for odd nucleon systems is significantly improved by employing a novel technique to accurately identify the block containing the ground state in odd configurations. 
To ensure the high numerical stability, the Kahan compensation algorithm is employed. The current version of the EP code can efficiently expand the computational space to handle up to 26 doubly folded (deformed) single-particle levels and 26 nucleons on a standard desktop computer in approximately $10^2$ seconds with double precision. With sufficient computational resources, the code can process up to 63 deformed single-particle levels, which can accomodate from 1 to 63 nucleon pairs. The EP v1.0 code is also designed for future extensions, including the finite-temperature and parallel computations.

%\end{linenumbers}
%\end{abstract}
%\begin{keyword}
%Nuclear pairing; Exact pairing solution (EP), Fortran, Sparse matrix, binary representation.

%\end{keyword}

%\end{frontmatter}

% \linenumbers

% All CPiP articles must contain the following
% PROGRAM SUMMARY.

{\bf PROGRAM SUMMARY}
  %Delete as appropriate.

\begin{small}
\noindent
{\em Program Title:} EP (v1.0)                                       
\\
{\em CPC Library link to program files:} (to be added by Technical Editor) 
\\
{\em Developer's repository link:} \url{https://github.com/ifas-mathphys/epcode_v01}
\\
{\em Code Ocean capsule:} (to be added by Technical Editor)
\\
{\em Licensing provisions:} GPLv3  
\\
{\em Programming language:} Fortran                                  
\\
{\em Supplementary material:}                                 
\\
  % Fill in if necessary, otherwise leave out.
{\em Nature of problem:}
\\
Nature of problem: Exact computation of eigenvalues and eigenvectors via direct diagonalization of the general nuclear pairing Hamiltonian for both odd and even nucleon configurations at zero temperature, ensuring high accuracy, enhanced numerical stability, and reduced computational time.
\\
{\em Solution method:}\\
Using the quasispin representation within the SU(2) algebra, we construct the pair-exchange matrix of the pairing Hamiltonian. To expedite the construction of the pairing matrix, we employ the binary representation to encode the information of paired states represented by matrix elements, while the reduction of computational time is achieved through the implementation of a sparse matrix diagonalization algorithm. An improved hash function, inspired by Ref. [1], is used to efficiently retrieve the indices of the pairing matrix, thereby speeding up its construction. A technique for determining the ground-state block in odd-nucleon configurations that significantly reduces the computation time is proposed. To ensure the high error stability, the Kahan summation algorithm is used. These techniques enhance the computational scale with a larger single-particle energy spectrum and reduce the computational time.
\\
{\em Additional comments including restrictions and unusual features:}
\\
  The present EP code is designed to accommodate a deformed single-particle spectrum, where each single-particle level can only be occupied by a maximum of two nucleons. For users intending to work with a spherical spectrum, it is necessary to partition the spherical single-particle levels into deformed shells for input purposes. It is important to note that this version does not include an extension of the EP code to handle finite-temperature conditions. The typical runtime of the code is heavily dependent on factors such as the number of nucleons provided as input, the chosen truncated space (with a maximum of 63 levels), 
the number of particles (with a maximum of 2$ \times $63 nucleons), the compilers (Gfortran/Ifort), and the configuration of the computer being used. Consequently, the execution time can range from less than one second to several days.
\\

\end{small}

\end{abstract}


\end{frontmatter}

\section{Introduction}
\label{Intro}
Pairing correlation is known to have a significant impact on the description of nuclear properties, ranging from nuclear structure to reactions \cite{Brink,Ring}. Drawing on the Bardeen-Cooper-Schrieffer (BCS) theory for superconducting electron systems \cite{BCS}, nucleons in a nucleus can form pairs with opposite spins at the same single-particle level to minimize the total binding energy of the system. This coupling mechanism is referred to as nuclear pairing correlation. Numerous calculations demonstrate the suitability of the proposed pairing model for nuclei \cite{Brink}. The inclusion of pairing correlations is essential to realistically describe nuclear ground-state properties such as binding energies, nucleon separation energies, and root-mean-square radii \cite{Brink}. 

Prominent methods to approach the nuclear pairing problem are the BCS theory, the Hartree-Fock Bogoliubov (HFB) method \cite{HFB}, and the exact solution of the pairing Hamiltonian \cite{Brink, Ring, Suhonen}. The first two methods (BCS and HFB) were applied very successfully to describe several nuclear structure properties. However, the particle number is not conserved within those methods unless projection techniques are applied prior to the application of the variational principle \cite{Ring}. At finite temperature, those methods predict the collapse of pairing gap at a critical temperature. This collapse, however, does not occur in finite nuclear systems due to the contribution of strong thermal and quantal fluctuations, as demonstrated by numerous theoretical and experimental studies \cite{HungRep}. These limitations can be overcome by using exact particle-number projection techniques \cite{PNP}. One candidate is the Lipkin-Nogami method \cite{LN}, which is both simple and easy to implement \cite{HungC81}. Practically, the pairing problem was exactly solved by the so-called Richardson method \cite{Brink,Ric1,Ric2}. This method was proposed based on solving a set of nonlinear equations, the so-call Richardson equations, to obtain the exact eigenstates of pairs in a given single-particle configuration. Solving the Richardson equations involves the use of intricate calculations, including those tied to infinite-dimensional algebras \cite{Pan}. A noteworthy point is that the nonlinear Richardson equations are valid only for a constant pairing strength $ G $, not for a pairing matrix $ G_{jj'}$. Therefore, instead of solving complex Richardson equations, recently, Volya \textit{et al.} developed a method, termed exact pairing (EP), to directly diagonalize the nuclear pairing Hamiltonian through its representation based on SU(2) quasispin algebras \cite{Volya}. This method, which is applicable for both constant $ G $ and matrix $ G_{jj'}$, is more practical for obtaining the exact solutions of the pairing Hamiltonian.

The EP solution has showcased its significance and utility in numerous pivotal studies since its inception. These studies encompass a broad spectrum, ranging from nuclear structure and nuclear reactions to nuclear astrophysics \cite{HungRep,Le,Zele}. The ability to exactly and easily obtain the eigenvalues and eigenvectors through the EP presents a substantial advantage in constructing nuclear pairing states at both zero and finite temperatures, overcoming the limitations of the conventional BCS. Consequently, this approach enables the exploration of nuclear pairing phase transitions at finite temperatures, as well as critical quantities pertaining to nuclear reactions, including nuclear level density (NLD) and radiative strength function (RSF) \cite{HungRep,Hung2017}. 

However, the shell-model-like pairing matrix within the EP grows rapidly in dimension as the number of single-particle configurations in the system increases, leading to significant computational demands in both resources and time. Traditional methods for directly diagonalizing the EP matrix become computationally prohibitive on a typical personal computer when the system size reaches 16 particles and 16 levels or more \cite{Hung2009}. Recently, a powerful program called PairDiag \cite{PairDiag} has introduced an innovative and efficient approach to constructing and diagonalizing the EP matrices. In particular, it represents the basis vectors of the EP matrix in a binary form, which accelerates the matrix construction process. Additionally, the iterative Lanczos algorithm was employed to compute the eigenvalues and eigenvectors of the EP matrix. Together, these techniques enable exact solutions to the pairing problem with expanded computational capacity and enhanced speed. However, PairDiag encounters two primary challenges that limit its computational efficiency. First, instead of storing the entire EP matrix for repeated use, PairDiag recalculates its elements each time they are accessed. Additionally, the code does not take advantage of the EP matrix's inherent symmetry and sparsity. Second, in calculating the occupation numbers, compensated summation is required to ensure numerical accuracy for the sums with large and variable dimensions. The Kahan summation algorithm \cite{Higham}, which addresses error compensation issues, has not been implemented in PairDiag, leading to unstable errors as computational space expands. An alternative approach for tackling the EP problem involving a large number of particles is the application of neural network quantum states through the variational Monte Carlo method \cite{Rigo}, which falls beyond the scope of the present paper. 

Inspired by the work of the PairDiag code, together with the usefulness of the EP problem, this paper introduces an accessible and robust program, allowing users to compute the EP solutions for deformed configurations at zero temperature. The EP v0.1 version we provide represents a significant upgrade from the conventional EP calculation code that has been previously used, though unpublished, in many of our earlier works (see, for example, \cite{HungRep,Hung2009} and references therein). This version addresses the limitations in the EP matrix construction and diagonalization in PairDiag as mentioned above. By improving the way to build the pairing sparse matrix through our fast construct algorithm, the current EP code version can expand the computational space up to 26 (deformed) single-particle levels with 1 to  26 nucleon pairs in a typical personal computer with 16 gigabytes (GB) of random access memory (RAM). Theoretically, this version can be scaled up to 63 single-particle levels (with 1 to 63 nucleon pairs), given sufficient hardware resources. Additionally, we propose enhanced algorithms for optimizing the EP matrix diagonalization along with the use of ARPACK and LAPACK packages, detailed in Section 3. We also introduce a technique for efficiently handling odd-nucleon configurations by exactly identifying and diagonalizing the ground-state blocks, which significantly accelerates the computational time. Moreover, the compensated summation problem is also treated by using the Kahan summation algorithm \cite{Higham} to ensure the high error stability and accuracy. Our program is also designed with flexibility, allowing for future extensions to spherical nuclear systems and finite-temperature conditions.

\section{Formalism}
\label{Forma}

In this section, we present the nuclear pairing Hamiltonian using the quasispin SU(2) representations. Subsequently, we elucidate the way to construct the pairing matrix elements from this Hamiltonian using the set of particle number operator $ N_j $ and seniority $ s_j $ along with the steps involved in configuring the computational space. The knowledge presented in this part is briefly presented based on the first work on the EP problem by Volya \textit{et al.,} in Ref. \cite{Volya}.

\subsection{Exact pairing solution at zero temperature}

We begin with a general pairing Hamiltonian, which describes  the pairing correlations in an atomic nucleus
  \begin{equation}
  \label{Ha0}
   \hat{H}
   =
   \sum_{jm} 
      \epsilon_{j} \, 
      a_{jm}^{\dagger} \, 
      a_{jm} 
      -
      \frac{1}{4} 
      \sum_{j j' m m'} 
         \textbf{G} ~ a_{jm}^{\dagger} \, 
         \widetilde{a}_{jm}^{\dagger} \, 
         \widetilde{a}_{j'm'} \,
         a_{j'm'} 
   .
  \end{equation}
In Eq. (\ref{Ha0}), the first term accounts for the nuclear mean field, while the subsequent term pertains to nuclear pairing effects. The physical parameters within these terms encompass the single-particle energy ($\epsilon_j$), the pairing strength ($\textbf{G}$), and the particle creation ($a_{jm}^{\dagger}$) and annihilation ($a_{jm}$) operators, where $ \widetilde{a}_{jm}=(-1)^{j-m}a_{j-m}$ denotes the time-reversal operator. The notation $jm$ designates a nucleon moving on the $j$th orbital with projections of $\pm m$. The pair degeneracies for such orbitals are given by $\Omega_j = j + 1/2$. It should be noted that $\textbf{G}$ can be a constant $G$ or can be a matrix $G_{jj'}$ describing the nucleon-pair-scattering strength between the $j$-th and $j'$-th orbitals or vice versa ($G_{jj'} = G_{j'j}$) \cite{Volya}. The matrix $G_{jj'}$ can be derived from the two-body interaction for monopole pairing $V_0$ as follows \cite{Suhonen,Zele}
\begin{equation}
\label{Vjj}
G(j,j') = \frac{2 V_0(jj,j'j')}{\sqrt{(2j+1)(2j'+1)}} ~~.
\end{equation}

The existence of quasispin symmetry in a paired system \cite{Racal} allows us to use the quasispin algebra to describe the quantum property of pairing correlations by introducing the partial quasispin operators as
\begin{eqnarray}
   \hat{L}_j^{-} 
&=& 
   \sum_m { \tilde{a}_{jm} \, a_{jm} }
,
\label{L_j-} 
\\ 
   \hat{L}_j^{+} 
&=&  
   ( \hat{L}_j^{-} )^\dagger 
= 
   \sum_m { a_{jm}^\dagger \, \tilde{a}_{jm}^\dagger }
, 
\label{L_j+} 
\\
   \hat{L}_j^z 
&=& 
   \frac{1}{2}
   \sum_m { ( a_{jm}^\dagger \, a_{jm} - \frac{1}{2} ) }
=
   \frac{1}{2} 
   ( \hat{N}_j - \Omega_j )
.
\label{L_j^z}
\end{eqnarray}
These quasispin operators follow the commutative property of the SU(2) algebra of angular momentum, namely 
\begin{equation}
   [ \hat{L}_j^+, \hat{L}_{j'}^- ] 
   = 
   2 \delta_{jj'} \hat{L}_j^z~
, 
\quad 
   [ \hat{L}_j^z, \hat{L}_{j'}^+ ] 
   = 
   \delta_{jj'}  \hat{L}_j^+~
, 
\quad 
   [ \hat{L}_j^z, \hat{L}_{j'}^- ] 
   = 
   - \delta_{jj'} \hat{L}_j^-~
.
\end{equation}
Using the quasispin operators, the Hamiltonian (\ref{Ha0}) can be rewritten as
\begin{eqnarray}
   \hat{H}
=
   \sum_j 
   { 
      \epsilon_j \, \Omega_j 
   } 
+ 
   2
   \sum_j 
   { 
      \epsilon_j \, \hat{L}_j^z 
   } 
- 
%% \textbf{G}_{jj'}
   \sum_{jj'} 
   \textbf{G} \, 
   {
      \hat{L}_j^+ \, \hat{L}_{j'}^{-}
   }
.
\label{HaEP}
\end{eqnarray}
For each $ \hat{L}_j $, it can be shown that 
$
   \hat{L}_j^2 = \hat{L}_j^+ \hat{L}_j^- - \hat{L}_j^z + (\hat{L}_j^z)^2 
$ 
commutes with $\hat{H}$. Therefore, $ L_j $ is also a good quantum number and makes the problem much easier to handle. The quasispin $ L_j $ and its projection $ L^z_j $ can be expressed in terms of the particle number $ N_j $ (occupancy) and unpaired particle $ s_j $ (seniority), as follows
\begin{equation}
   L_j 
= 
   \frac{1}{2} ( \Omega_j - s_j )
,
\quad 
   L_j^z 
= 
   \frac{1}{2} ( N_j - \Omega_j )
.
\label{Ns}
\end{equation}
Based on these expressions, the representation $ \vert L_j, L^z_j \rangle $ can be replaced by the set of basic states 
$ \vert \{ s_j \} , \{ N_j \} \rangle $, 
where 
$
   \{ s_j \}
=
   \{ s_1, \ldots, s_j, \ldots, s_{\Omega} \}
$
and 
$
   \{ N_j \} 
=
   \{ N_1, \ldots, N_j, \ldots, N_{\Omega} \}
$,
with the constraints $ s_j \leqslant N_j \leqslant 2\Omega_j-s_j $ and $ s_j \leqslant \Omega_j $. $ N= \sum_j N_j $ and $ s=\sum_j s_j $ are the total number of particles and unpaired particles, respectively.

Using the property of angular momentum operators, namely
\begin{equation}
   \hat{L_j}^{\pm} 
   \left| L_j , L_j^z \right \rangle 
= 
   \sqrt{ 
      (L_j \mp L_j^z)(L_j \pm L_j^z + 1)
   }
   \left| L_j, L_j^z \pm 1 \right \rangle
,
\end{equation}
the matrix elements of the pairing Hamitonian (\ref{HaEP}) can be constructed in the presentation of basic states 
$ \vert \{ s_j \} , \{ N_j \} \rangle $. 
The diagonal elements of the pairing matrix are given as
\begin{equation}
\resizebox{ .89 \textwidth}{!} 
{$
   \left\langle 
      \left\{ s_j \right\}
      ,
      \left\{ N_j \right\}
      \vert \; 
      \hat{H} 
      \; \vert
      \left\{ s_j \right\}
   ,
      \left\{ N_j \right\}
   \right\rangle 
= 
   \sum_j
   \left( 
      \epsilon_j N_j 
      - 
      \frac{ 1 }{ 4 } 
%%    \textbf{G}_{jj} 
      G_{jj} 
      \left( N_j - s_j \right) 
      \left( 2 \Omega_j - s_j - N_j + 2 \right) 
   \right)
,
$}
\label{diag}
\end{equation}
whereas, the off-diagonal elements, which describe the pair transfers, are given as
\begin{equation}
\resizebox{.89 \textwidth}{!} 
{$
   \begin{split}
   &
      \left\langle 
         \left\{ s_j \right\} ,
         \left\{ \ldots , N_j + 2 , \ldots,  N_{j'}-2 , \ldots \right\} 
         \vert \;  \hat{H}  \; \vert 
         \left\{ s_j \right\} , 
         \left\{ \ldots, N_j , \ldots, N_{j'} , \ldots \right\}
      \right\rangle 
   \\
   &=
      -
      \frac{ 1 }{ 4 }
%%    \textbf{G}_{jj'}
      G_{jj'}
      \left[ 
         \left( N_{j'} - s_{j'} \right) 
         \left( 2\Omega_{j'} - s_{j'}- N_{j'} + 2 \right) 
         \left( 2\Omega_{j} - s_{j} - N_{j} \right)
         \left( N_{j} - s_{j} + 2 \right)
      \right] ^{1/2}
.
\end{split}
$}
\label{offdiag}
\end{equation}
Direct diagonalization of the above pairing matrix allows us to obtain  the EP solution including the eigenvalues $ {\cal E}_s $ and eigenvectors $ \vert s \rangle = C_k^{(s)} \vert k \rangle $ corresponding to each total seniority $s$ state, where
$ 
   \vert k \rangle
=
   \vert \{ s_j \} , \{ N_j \} \rangle 
$ 
and 
$ \sum_k  (C_k^{(s)})^2 =1  $. 
The degeneracy of the $s$-th eigenstate is given as
\begin{eqnarray}
   d_s 
= 
   \prod_{j} 
   \left[
      \frac{ (2\Omega _j)! }{ s_j! \, ( 2\Omega - s_j )! }
      -
      \frac{ (2\Omega _j)! }{ (s_j-2)! \, (2\Omega - s_j + 2)! }  
   \right]
. 
\label{dS}
\end{eqnarray}   
The state-dependent exact occupation number $f_j^{(s)}$, which represents the probability of a particle occupying the $j$-th level, is calculated based on the average weight of partial occupancy $N_j^{(k)}$ over each $\vert k \rangle$ state, namely
\begin{eqnarray}
   f_j^{(s)} 
= 
   \frac{
      \displaystyle 
      \sum_k N_j^{(k)} \left( C_k^{(s)} \right)^2
   }{
      \displaystyle 
      \sum_k \left( C_k^{(s)} \right)^2
   }
=
   \displaystyle 
   \sum_k N_j^{(k)} \left( C_k^{(s)} \right)^2
,
\label{fEP0}
\end{eqnarray}

with 

\begin{equation}
\sum_k \left( C_k^{(s)} \right)^2 = 1.
\end{equation}

\subsection{Characteristics of the computational configuration}

Obtaining the EP solution primarily hinges on the construction and diagonalization of the pairing matrix. Due to the limitations of the truncated space, as the dimension of the pairing matrix in the EP problem increases rapidly with increasing the number of particles and single-particle levels, previous EP calculations were only capable of solving configurations with a number of deformed levels and particles below 16 while running efficiently on a common personal computer \cite{HungC82}. When performing the EP calculations, the choice of truncated space including the numbers of deformed single-particle levels $ \Omega $ and nucleons $ N $ can be tailored to the user's requirements. The single-particle energy $ \epsilon_j $ corresponding to the chosen levels are typically extracted from the single-particle energy spectrum obtained using the shell model or mean-field approach. It is essential to note that $ N $ and $ \Omega $ can be either even or odd, and these two values do not need to match ($N \leq 2\Omega$). The single-particle levels can take on various types, including deformed or spherical orbits. The pairing strength, denoted as $\textbf{G}$, can be a constant $G$ or a matrix $G_{jj'}$ depending on the user's requirements.

\section{Methodology of implementation}
\label{Meim}

\subsection{Evaluate the EP matrix}
\begin{figure}[H]
\begin{center}
\includegraphics[width=1\textwidth]{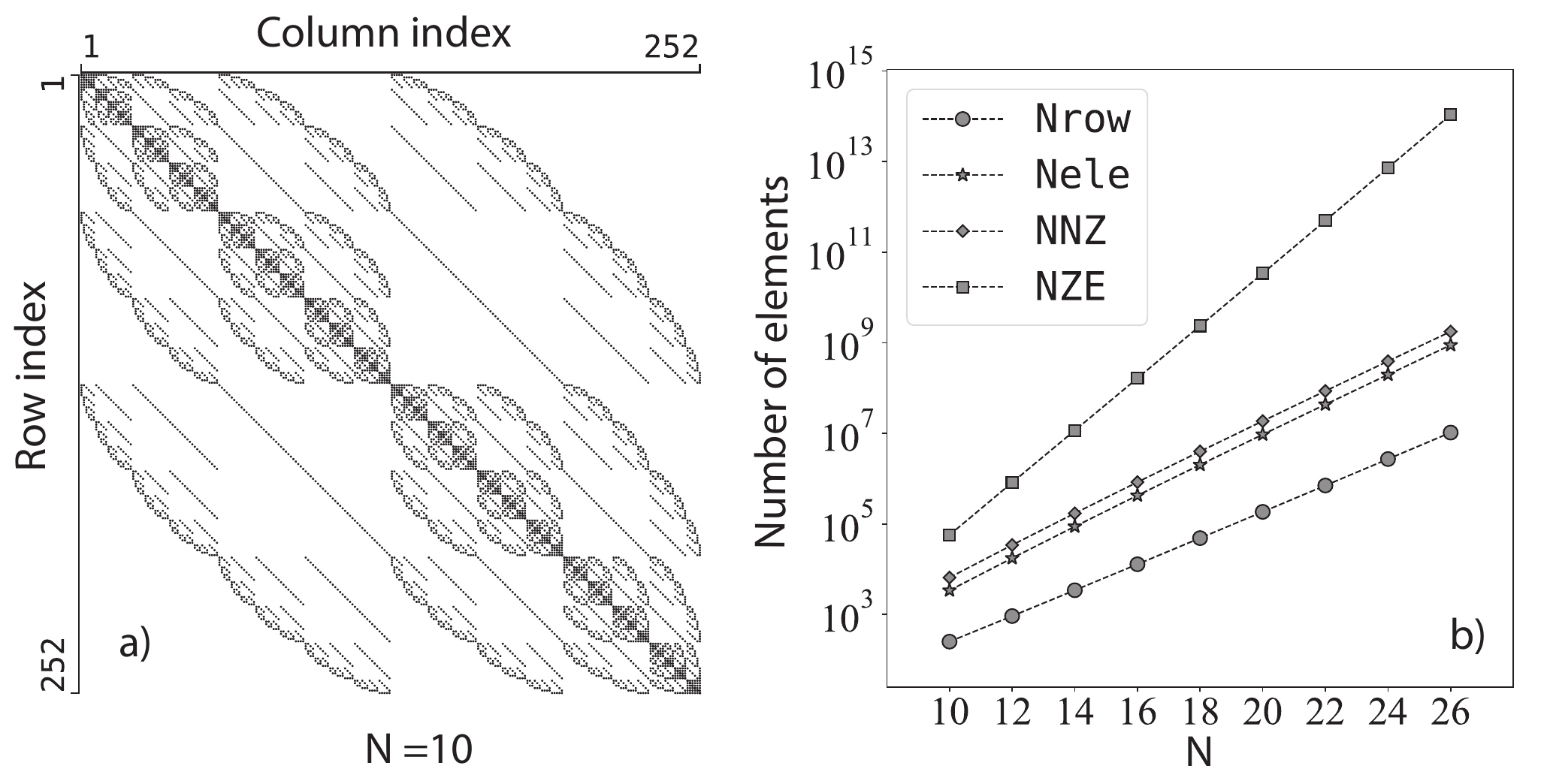}
\end{center}
\caption{%
Sparse nature of the pairing matrix of $\hat{H}$. The left panel (a) 
shows the non-zero elements (solid dots) in the system of $N=\Omega=10$. The right panel (b) shows the log10 scale of the number of matrix rows ({\tt Nrow}) and elements. Wherein, the dimension of $\hat{H}$ is $ {\tt Nrow} \times {\tt Nrow}$. {\tt Nele} and {\tt NNZ} are, respectively, the numbers of non-zero elements of the upper triangular part of $\hat{H}$ and that of the full $ \hat{H} $, respectively, while {\tt NZE} is the number of zeros in $ \hat{H} $ versus $N=\Omega$, for $ 10 \le \Omega \le 26  $ and $\Omega $ even.
}
\label{fig1}
\end{figure}

The pairing matrix of $\hat{H}$, a symmetric real matrix with real eigenvalues and eigenfunctions, is constructed from the diagonal (\ref{diag}) and off-diagonal (\ref{offdiag}) components. At zero temperature ($T=0$) without pair breaking, nucleon pairs occupy single-particle levels and scatter to unoccupied ones. For a system with $\Omega$ deformed single-particle levels and $N$ nucleons, where $N \leq 2\Omega$, all possible configurations must be considered. At $T=0$, this yields $C^\Omega_{N/2}$ configurations for even $N$ and $\Omega C^{\Omega-1}_{(N-1)/2}$ configurations for odd $N$, due to the Pauli blocking by the unpaired nucleon. Pairing interactions produce the non-zero diagonal matrix elements, while the off-diagonal elements are non-zero only when the scattering of a pair occurs between levels. For instance, the pairing matrix elements of a system with $\Omega = 10$ and $N = 10$ are illustrated in Fig. \ref{fig1}(a).

The pairing matrix, as shown in Fig. \ref{fig1}(a), is sparse due to numerous zero elements. It is evident that configurations with $\Omega = N$ have the highest density of non-zero elements, whereas cases with $\Omega \neq N$ contain more zero elements. Hence, we focus solely on the case $\Omega = N$ for illustration. Fig. \ref{fig1}(b) shows the evaluation of non-zero elements for $N = \Omega$ configurations, with $10 \le \Omega \le 26$ and $\Omega$ even. As $N$ increases, more zero elements appear, enhancing sparsity. Diagonalizing sparse matrices as regular matrices, as in prior works \cite{HungRep,Volya,Zele,Hung2009,HungC82}, increases the computational cost and limits the scalability. Therefore, this work applies sparse matrix algorithms to enhance the efficiency of finding the EP solutions.

%%%%%%%%%%%%%%%%%%%%%%
%  DO NOT DELETE THIS if we still need it 
%
%  f01_Hmat_N10.pdf     -> Hmat.pdf
%                          Hmat_N10.txt + x_Hmat.py 
%
%  f02_Hmat_Sizes.pdf   -> Hmat_Sizes.pdf
%                          z_Siz_gfortran_even.txt + x_Hmat_Sizes.py
%
%
%%%%%%%%%%%%%%%%%%%%%%%%%

\subsection{Idea to fast construct and diagonalize the EP matrix}

\subsubsection{Binary representation}

\begin{figure}[!h]
\includegraphics[width=1\textwidth]{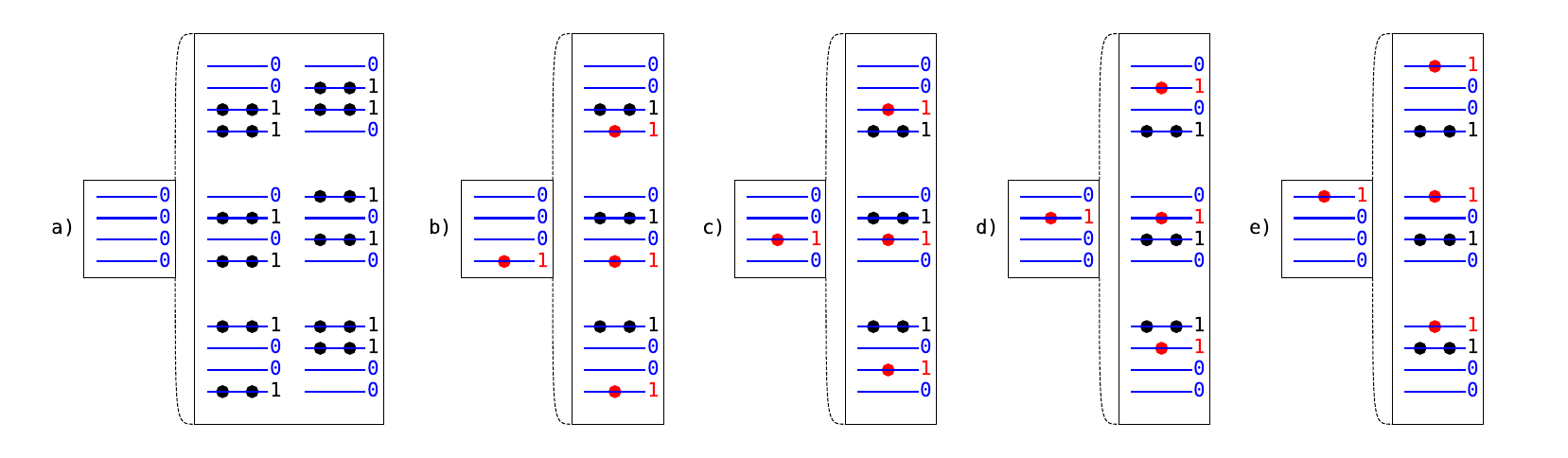}
\caption{\label{fig2}% 
      Illustration of all possible configurations of an even system with $N=\Omega =4$ (a), 
      and an odd system with $\Omega =4$ and $N=3$ (b)-(e) in the binary representation.
}
\end{figure}

The burden of computing the EP solution primarily comes from two factors: 
a) constructing the pairing matrix and 
b) diagonalizing it. 
To optimize the first factor, the idea of binarizing the nucleon states is used \cite{PairDiag}. 
A single-particle state in the occupation-number representation can be expressed as a binary digit (bit), with 1 indicating occupation and 0 indicating non-occupation. 
A possible state of the system is thus represented by sequences of such bits. 
In the current version of the EP code, we employ a space with 64 bits, 
where the first bit signifies the sign and the subsequent 63 bits encode the states of the system. 
In the context of zero-seniority pairing $s=0$, 
a pair of nucleons and its occupancy on a level can be represented by a single bit. 
It is important to note that each level now accommodates only two nucleons, 
corresponding to just one pair. 
In odd configurations ($s=1$), the unpaired nucleon occupies a level, blocking it and leaving the remaining levels available for paired nucleons.
Fig. \ref{fig2} illustrates all possible configurations for an even $N=\Omega=4$ and an odd $N=3$, $\Omega=4$ systems in the binary representation.

%%%%%%%%%%%%%%%%%%%%%%
%% Figure2
%%
%%%%%%%%%%%%%%%%%%%%%%%

Fig. \ref{fig2} shows that for $s=0$, there is one pairing matrix, but for $s=1$, the matrix splits into smaller blocks with smaller dimensions (e.g., 4 blocks in Fig. \ref{fig2}(b)-(e)) due to the unpaired particle. The unpaired particles do not affect the off-diagonal structure of $\hat{H}$ across the blocks for a given $s$. Therefore, in the first step, $\hat{H}$ can be constructed based on the occupied states of nucleon pairs without immediate concern for unpaired particles. This approach is highly effective for developing the EP code at finite temperature and such an occupied state can be called the post-occupied state (POS). For instance, in Fig. \ref{fig2}(b)-(e), excluding the unpaired particle leaves three positions available for a pair, yielding three POSs: $\vert 100 \rangle$, $\vert 010 \rangle$, and $\vert 001 \rangle$. The off-diagonal elements, identical across the four blocks, are calculated one time and applied to all. The unpaired particles are only considered for the main diagonal of $\hat{H}$ (Section \ref{H-jah-aah}). This process is implemented through the \textbf{gens} subroutine in the code, utilizing bitwise algorithms. A pseudocode outlining \textbf{gens} is provided in {\bf Alg.} \ref{alg-gens}.

\begin{center}
\begin{algorithm}[H]
\label{alg-gens}
\caption{%
Procedure 
\textbf{gens} 
generates a set of integers representing all possible POSs 
in ascending bit order and return the result to 
$ \{ S_m \}_{ m = \overline{1,L} } $.
Herein 
$ p $ is the total number of 1-bits in a state,
$ p \le \omega $, 
$ \omega $ is the number of active bits in $S_m$ 
(where $S_m$ is treated as an integer) 
and also represents the number of available levels in $S_m$ 
(which in this context is treated as a state),
$ \omega = \Omega - s $,
$ \Omega$ 
is the total number of levels,
$s$ is the seniority,
$ L $ 
is the length of $\{ S_m \}$,
i.e., 
$
   L = C^{\omega}_{p} 
$, 
and
IOR and 
IAND 
are Fortran bit manipulation intrinsic functions.}
\Input{ $ p, L $ }
\Output{ $ \{ S_m \}_{ m = \overline{1,L} } $ }
\Local{ $ m, k, j_0, j_1, j_2, j_3 $ } \;  
   $S_1 \gets 2^p - 1$ \;  
   \For{ 
      $ m \gets 2 $ \KwTo $ L $
   }{
      $ j_0 \gets  S_{m-1} $ \; 
      $ j_1 \gets {\rm IOR }( j_0, j_0-1 ) $ \; 
%\tcp*{turn on the trailing 0-bits of $j_0$}
      $ j_2 \gets j_0 - {\rm IAND } ( j_1 + 1, j_0 ) $ \; 
%\tcp*{keep only the rightmost adjacent 1-bits of $j_0$}
      $ k   \gets  -1 $ \;
      \While { $ j_2 \neq 0 $ }{ 
         $ k   \gets  k + 1 $ ; 
         $ j_2 \gets  {\rm IAND} ( j_2, j_2 - 1 ) $ \; 
      } 
      $ S_m \gets j_1 + 2^{ k } $ \; 
   }
\end{algorithm}
\end{center}

In the initial phase of \textbf{gens}, the first POS is allocated using the $p$ rightmost adjacent 1-bits in the binary representation.
Once this state is determined, the next task is to identify and organize all remaining states of the system in ascending order of integers. In our current code, the \textbf{gens} subroutine is inspired by the \textbf{Next\_State} subroutine in the PairDiag code
\cite{PairDiag}. However, in \textbf{gens}, we implement fully bitwise intrinsic functions to enhance the execution speed. Further details on the bitwise calculations can be found in Ref. \cite{hack}. 
The \textbf{gens} subroutine also works well with seniorities greater than one. Its current design proves particularly advantageous for extending to finite-temperature calculations, where seniority $ s > 1 $ primarily arises from thermal effects.

\begin{center}
\begin{algorithm}[H]
\label{alg-npairing}
\caption{%
Procedure 
\textbf{npairing} 
counts  
from 
$ S $ 
the number of its pair-scattering states, 
which have larger in value compared to $ S $, 
and return the result to $ \varkappa $. 
Herein 
$ S $ is a POS,
IAND and ISHFT are Fortran intrinsic functions, and
$ \omega $ 
is defined in {\bf Alg.} \ref{alg-gens}.
}
\Input{ $ S$, $\omega $ }
\Output{ $ \varkappa $ }
\Local{ $ t, j, i  $ } \;  
   $ \varkappa \gets 0 $ ;  $ t \gets S $ ; $ j \gets 0 $ ;  $ i \gets 0 $ \; 
   \While{ 
      $ t \neq 0 $  
   }{
      \eIf{
         ${\rm IAND} ( t, 1 )  \neq 0 $
      }{
         $j \gets j + 1 $  \; 
      }{
         $\varkappa \gets \varkappa + j $  \; 
      }
      $ t \gets {\rm ISHFT} ( t, -1 ) $ ;  
      $ i \gets i + 1  $ \;
   }
   $ \varkappa \gets  \varkappa  + (\omega - i ) j $ \; 
\end{algorithm}
\end{center}

After constructing a configuration with all possible states in the binary representation, the next step is to determine the locations where the off-diagonal pair scattering occurs. 
{\bf Alg.} 
\ref{alg-npairing} demonstrates the \textbf{npairing} 
procedure, which counts the total number of pair-scattering states for a given POS in ascending bit order.
The \textbf{npairing} procedure determines the exact amount of memory required to allocate an array for the scattering states. Next, {\bf Alg.} 
\ref{alg-genp} describes the \textbf{genp} 
procedure that generates precisely from a given POS a set of all its possible
pair-scattering states, which are larger in value compared to this POS, i.e.,
only the upper triangular matrix is considered.

\begin{center}
\begin{algorithm}[H]
\label{alg-genp}
\caption{%
Procedure 
\textbf{genp} 
generates from 
$ S $ 
all its pair-scattering states, which are larger in value, 
and return the result to 
$
   \{ R_k \}_{ k = \overline{  1 , \varkappa } } 
$. 
Here
$ S $ is a POS,
$ \varkappa $ 
is the counter determined in {\bf Alg.} \ref{alg-npairing}, where
%% the length of   
%% $ R $ 
%% is not larger than 
%% $
%%    p ( \omega  - p ) 
%% $, 
%% i.e., 
$ \varkappa \le p ( \omega  - p ) $, and
$ \omega $ and $ p $
are defined in {\bf Alg.} \ref{alg-gens}.
}
\Input{ $ \omega, p, S $ }
\Output{ $ \varkappa $,  $ \{ R_k \}_{ k = \overline{  1 , \varkappa  } }  $ }
\Local{ $t$, $l$, $j$, $i$, $ \{ a_l \}_{ l = \overline{ 1, \omega } } $ } \; 
   $ t \gets S $ ; 
   $ \varkappa \gets 0 $ ;
   $ j \gets 0 $ ;
   $ i \gets 0 $ ;
   $ a_l \gets 0 $, $\forall l = \overline{ 1, \omega } $ \;
   \While{ 
      $ i < \omega $ 
   }{
      \eIf{
         $ {\rm IAND}( t, 1 ) \neq 0 $
      }{
         $ j \gets j + 1 $ ; 
         $ a_j \gets i $ \; 
      }{
         \For{ 
            $ l \gets j $  \KwTo $ 1 $  \KwBy $-1$ 
         }{
            $ \varkappa \gets \varkappa + 1 $ \;
            $ R_{\varkappa} \gets S $ \; 
            $ R_{\varkappa} \gets {\rm IBSET} ( R_{\varkappa}, i ) $ \; 
            $ R_{\varkappa} \gets {\rm IBCLR} ( R_{\varkappa}, a_l ) $ \; 
         }
      }
      $ t \gets {\rm ISHFT} ( t, -1 ) $ ; 
      $ i \gets i + 1  $ \; 
   }
\end{algorithm}
\end{center}

\begin{center}
\begin{algorithm}[H]
\label{alg-hashs}
\caption{%
Procedure 
\textbf{hashs} 
retrieves the global index of $ S $ 
and return the result to $ n $. 
Here
$ S $ is a POS,  
$ \omega  $ and $p$
are defined in {\bf Alg.} \ref{alg-gens},
$
   C^{ m }_{k} 
=
   ( m! ) / ( k! (m-k)! )
$, 
and 
$
   C^{m}_{k} 
=
   0
$
as $ m < k$.
}
\Input{ $ \omega, p, S $ }
\Output{ $ n $ }
\Local{ $ t $, $i_l$, $i_h$ } \; 
   $ t   \gets S $ ;  
   $ n \gets 1 $ ; 
   $ i_l \gets 0 $ ; 
   $ i_h \gets 0 $ \; 
   \While{ 
      $ t \neq 0 $ 
   }{
      \If{
         $ {\rm IAND} ( t, 1 ) \neq  0 $ 
      }{
         $ i_h \gets i_h + 1 $ ; 
         $ n \gets n + C^{i_l}_{i_h}  $ \;
         \If{  
            $i_h = p$ 
         }{  
            exit {\bf while} \; 
         } 
         $ t = {\rm ISHFT} ( t, -1 )$ \;
         $ i_l = i_l + 1 $ \;
      }
   }
\end{algorithm}
\end{center}

The remaining task is to find the indices of non-zero elements of
$\hat{H}$ 
corresponding to the scattering states obtained from \textbf{gens}.
{\bf Alg.}
\ref{alg-hashs}
shows the 
\textbf{hashs} 
procedure that retrieves the global index of a given POS.  
This procedure is inspired by
the {\bf Hash\_State}  
subroutine in the PairDiag code
\cite{PairDiag}.
The roles of 
\textbf{npairing}, \textbf{genp}, and \textbf{hashs} 
procedures will be discussed in more detail in Section \ref{store}

\begin{center}
\begin{algorithm}[H]
\label{alg-gind}
\caption{%
Procedure 
\textbf{gind} 
retrieves the index 
$ ( j, k ) $ 
of 
$ G_{ j k } $
in Eq. \eqref{HaEP} 
from 
$ S_1 $  
and 
$ S_2 $.
Here 
$ S_1 $  
and 
$ S_2 $ are the POS and its pair-scattering state, respectively. 
IEOR, IAND, IOR, and POPCNT
are Fortran bit manipulation intrinsic functions.
}
\Input{ $ S_1, S_2 $ }
\Output{ $ j, k $ }
\Local{ $ x, y, z $ } \; 
   $x  = {\rm IEOR} ( S_1, S_2 )$ \;
   $z = {\rm IAND} ( x, x-1 )$ \;
   $y = x - z $ \;
   $k = {\rm POPCNT} ( {\rm IOR} ( z, z - 1 ) ) $  \;
   $j = {\rm POPCNT} ( {\rm IOR} ( y, y - 1 ) )$ \;
\end{algorithm}
\end{center}

If the pairing strength 
 \textbf{G}
is used as a matrix in (\ref{Vjj}), the values of $ G_{ j k } $ 
are required where the scattering of a pair occurs. In this case, 
{\bf Alg.} 
\ref{alg-gind}
provides the 
\textbf{gind} 
procedure that determines the index 
$ ( j, k ) $ 
of 
$ G_{ j k } $
in \eqref{HaEP} 
from a given POS and its pair-scattering state.

\begin{center}
\begin{algorithm}[H]
\label{alg-dec2bin}
\caption{%
Procedure 
\textbf{deccon} 
extracts from 
$ S $ 
an array $b$ 
of 0s and 1s as decimal integers,
i.e., 
the $\omega$ 
active rightmost adjacent bits in the binary representation of $S$. Here 
$S $ is a POS and 
$ \omega $ 
is defined in {\bf Alg.} \ref{alg-gens}.  
IAND and ISHFT are the Fortran bit manipulation intrinsic functions.
}
\Input{ $ S, \omega $ }
\Output{ $ \{ b_i \}_{ i = \overline{1, \omega } } $ }
\Local{ $ t, i, l $ } \; 
   $ t \gets S $ \;  
   \For{ 
      $ i \gets 1$   \KwTo  $\omega $  
   }{
      $ b_i = {\rm IAND} ( t, 1) $ ;
      $ t = {\rm ISHFT} ( t, -1) $  \;
      \If { 
         $ t = 0 $  
      }{
         exit {\bf for} \;
      }
   }
    \For{ 
      $ l \gets i+1 $   \KwTo  $ \omega $  
   }{
      $ b_l = 0 $ \; 
   }
\end{algorithm}
\end{center}

{\bf Alg.}
\ref{alg-dec2bin}
shows the 
\textbf{deccon} 
procedure that extracts from a given POS an array of 0s and 1s as decimal integers, 
which are all its active rightmost adjacent bits in the binary representation.
This procedure is primarily used for calculating the diagonal elements of 
$ \hat{H} $.

\subsubsection{Retrieving and storing pairing matrix}
\label{store}

Once all possible states of the system have been generated, 
it is crucial to investigate quickly all the pair-scattering states in the off-diagonal of 
$ \hat{H} $. 
This section serves for estimating the number of non-zero elements 
and their indices in $\hat{H}$. Since $ \hat{H} $ is symmetric and sparse, 
it is sufficient to store only its upper triangular
portion, i.e., $H_{ m, n } \neq 0 $ for $ n \ge m $, in a compressed format.

For convenience in explaining precisely how our procedures work, we make use of 
some mathematical notations. 
Let us denote a POS and its corresponding pair-scattering state by $S_m$ and $S_{m_k}$, respectively.
Here,  
the array of POS,
$\{ S_m \}_{m = \overline{ 1 , L} } $, 
is generated in advance by 
\textbf{gens} ({\bf Alg.} \ref{alg-gens}) 
and stored in a strictly ascending array of integers,
i.e., 
$ S_{j} > S_{k} $ 
if and only if  
$ j > k $, where
$L$ is defined in the {\bf Alg.} \ref{alg-gens}.
%% Moreover, in this version of EP, 
%% $S_k$ are 64-bit signed integers.
It is essential to emphasize that  
$m$ and $ m_k$
are the global indices of the states.

To demonstrate the structure of $\hat{H}$,
let us also denote by $I$ the set of these global indices, i.e.,
$
   I = 
   \{
      1, 2 , \ldots, L
   \}
$,
and let $ \bU_m $ denote the $m$-th column vector of the identity matrix
$\bI_L$,
for $ m = \overline{1,L} $.
Based on these assignments, we can rewrite Eqs. 
\eqref{offdiag} 
and 
\eqref{diag}
as
$
   H_{ m, m_k }
=
   \bU_m ^T \, 
   \hat{H} \,
   \bU_{m_k} 
$
for the non-zero off-diagonal entries of $\hat{H}$, 
and
$ 
   H_{ m, m }
=
   \bU_m^T \, \hat{H} \, \bU_{m}
$
for the elements in the main diagonal of $\hat{H}$, 
respectively,
where 
$
   m, m_k \in I 
$
and 
$
   m_k \neq m 
$.

   For each non-zero off-diagonal element
   $H_{m, m_k }$, 
   we aim to quickly calculate 
   the column indices $m_k \in I $ 
   on the $m$-th row such that $ m_k > m $,
   for every $m \in I$. 
   To serve this purpose, 
   the \textbf{npairing}, \textbf{genp}, and \textbf{hashs} procedures are used,
   as follows.

First, %to count the number of such non-zero elements
for every 
$m \in I$,
the \textbf{npairing} function
({\bf Alg.} \ref{alg-npairing}) 
counts quickly from $ S_m $ 
the total number (i.e., $\varkappa$)  
of all the pair-scattering states $\{ S_{m_k} \}_{ k = \overline{1, \varkappa} }$
such that 
$ 
   S_{m_k} > S_{m} 
$,
for $ k = \overline{1, \varkappa} $,
and 
$\{ m_k \}_{ k = \overline{1, \varkappa} } \subset I$.
This calculation serves only for the computer memory allocation.

Second, 
for every 
$m \in I$,
the 
\textbf{genp} procedure 
({\bf Alg.} \ref{alg-genp})
determines precisely from $ S_m $ 
all its pair-scattering states $ S_{m_k} $
such that 
$ 
   S_{m_k} > S_{m} 
$,
and
$
   \{ 
      S_{ m_k } 
   \}_{ k = \overline{1,\varkappa} }
$
is a strictly ascending array of integers, where 
$ \varkappa $ is also determined by \textbf{npairing}.
It should be noted that
$
   \{ 
      S_{ m_k } 
   \}_{ k = \overline{1,\varkappa} }
   \subset 
   \{ 
      S_{ m } 
   \}_{ m \in I }
$
and 
$
   \{ 
      m_k
   \}_{ k = \overline{1,\varkappa} }
   \subset 
   I
$. 
Since
$
   \{ 
      S_{ m_k } 
   \}_{ k = \overline{1,\varkappa} }
$
is a strictly ascending sequence, where
$ 
   S_m 
<
   S_{m_k}  
<
   S_{m_{k+1}}  
$ 
imply  
$
   m
   <    
   m_{k}
   < 
   m_{k+1} 
$.
Wherein, $m$ 
serves as the row index along the main diagonal, 
while $m_k$ represents the column indices in the $m$-th row.
Jointly, they play the primary role in the data structure for $\hat{H}$.
%Let us emphasize that the values of  
%$
%   \{ 
%      m_k
%   \}_{ k = \overline{1,\varkappa} }
%$
%are still unknown in the scope.

Last,
the  
 \textbf{hashs} function 
({\bf Alg.} \ref{alg-hashs})
retrieves the global index of 
$ S_{m_k} $, 
i.e., 
finding 
$  
   n \in I
$ 
uniquely such that 
$ 
   S_n = S_{m_k} 
$.
Therefore,
the column index $m_k$
of a non-zero off-diagonal element
$ H_{m, m_k }$ 
is determined,
for every 
$ m \in I $.
For example, let's consider the state $\vert 0010 \rangle$. The pair-scattering states corresponding to this are $\vert 0100 \rangle$ and $\vert 1000 \rangle$.
This implies that \textbf{npairing}($ \vert 0010 \rangle $) $ = b =2$. 
However, the state $\vert 0001 \rangle $, although it has a pair scattering from the initial state, 
is ignored because we only consider the elements of the upper triangular matrix, namely, from $j$ to $j'>j$.

%%%%%%%%%%%%%%%%%%%%%%
%%Figure4
%% \begin{figure}[!h]
%% \includegraphics[width=1\textwidth]{npairinggenp.jpg}
%% \caption{\label{fig4} %
%%    A pseudocode describes the state generation algorithm.
%% }
%% \end{figure}
%%
%%%%%%%%%%%%%%%%%%%%%%%

Once the number of non-zero elements is known, 
an array to store the non-zero entries of the upper triangular portion of the matrix is allocated. 
These non-zero elements are organized into separate segments of the array: 
one segment for the diagonal elements and another for the off-diagonal elements. 
This storage approach, known as the modified compressed sparse row (MSR) format \cite{MSR}, 
is particularly well-suited for large scale sparse matrix diagonalization procedures, 
as we will discuss in the following subsection.

In the \textbf{hashs} subroutine, it should be noted that we also use the hash function from Ref. \cite{PairDiag}, with a small improvement to determine the index of the matrix elements.
The hash function is given as
\begin{equation}
   f ( \vert \lbrace s \rbrace, \lbrace N_j \rbrace \rangle ) =  1 +  \sum_{i=1}^p   C^{O_i-1}_i ,
\end{equation}
where 
$p$ is the number of occupied orbits (the number of bit `1' in a state), and 
$O_i$ is the order of $i$th bit `1'. 
For instance, 
consider a state $\vert 0101 \rangle $ 
where there are two `1' bits ($p=2$), 
the first one is positioned at the 1$^{\text{st}}$ bit, and 
the second one is at the 3$^{\text{rd}}$ bit, counting from right to left. 
Thus, $O_1=1$ and $O_2 = 3 $, 
and the index of this state is 
$
   f(\vert 0101 \rangle) = 1 + C^{O_1-1}_1 + C^{O_2-1}_2 = 2
$. 
This function will stop as soon as it scans enough bit `1' of the input state. The \textbf{hashs} is used 
immediately after the \textbf{genp} subroutine 
to determine the location of non-zero matrix elements. 
The value of these elements will then be calculated and assigned through equations 
(\ref{diag}) and (\ref{offdiag}). 
After the non-zero elements are stored, 
the fast construction process of the pairing matrix is completed.

\subsubsection{Matrix Diagonalization}
\label{eigenpair}

For the main task of diagonalizing paring matrices, we consider two scenarios.
First, for the large $\hat{H}$ matrix of size $n \times n$,
where $n > n_0$ and $n_0 $ is a fixed integer constant (e.g., $n_0 = 100$), 
we treat it as a sparse matrix and apply the implicitly shifted QR algorithm combined with a 
$k$-step Arnoldi factorization 
to obtain the first few ($k$) smallest eigenvalues. Meanwhile, the corresponding eigenvectors are also obtained from the Schur basis of the $k$-dimensional eigenspace, and their orthogonal property is guaranteed in a working precision. This technique is known as the implicitly restarted Arnoldi method (IRAM). Since the method avoids performing the full iteration of the implicitly shifted QR algorithm, it requires less computer memory, making it suitable for large scale eigenpairing problems. In this scenario we use the DSAUPD subroutine in the ARPACK package \cite{arpack}. It is worth mentioning that the DSAUPD is designed with a reverse communication interface, meaning we do not have to pass the matrix as an argument; instead, we only need to provide an external procedure to perform matrix-vector multiplication. Therefore, we are free to choose a suitable data structure for the matrix $\hat{H}$. 
In the EP code, the modified compressed sparse row format (MSR) \cite{MSR} is applied to store the non-zero elements of $\hat{H}$.

Second, for $n \le  n_0$, 
we treat $\hat{H}$ as a dense matrix and fully apply the implicitly shifted 
QR algorithm.
In this scenario, 
the DSYEV subroutine in LAPACK package \cite{lapack} is utilized with the eigenvector option.
It is remarkable to note that the DSYEV depends on the DSTEQR,
which performs the implicitly shifted QR algorithm. 

%% In LAPACK subroutine DSYEV, to compute only eigenvalues, 
%% DSYEV calls DSYTRD and DSTEQR without eigenvector option. 
%% To compute both eigenvalues and eigenvectors, 
%% DSYEV calls DSYTRD, DORGTR, and DSTEQR with the eigenvector option. 

\subsubsection{%
Searching ground-state block in odd-system
}%
\label{odd-sys}

It is evident that odd configurations always result in 
   the partitioning of the pairing matrix into smaller blocks, depending on the initial position of the odd particle. 
Fig. \ref{fig2}(b)-(e) demonstrate the partitioning of the pairing matrix for the configuration 
($\Omega=4, N=3, s=1 $) 
into four blocks. 
To compute the ground-state eigenvalue of an odd configuration, 
we typically need to diagonalize all these blocks individually 
and then 
compare to select the smallest eigenvalue. 
However, if we can determine in advance which block contains this value, 
we can save significant computational time by diagonalizing only that block. 
In this work, 
we propose a straightforward approach to identify such a block 
   by calculating the effective energy of a state 
($ E_{\rm eff} $), 
defined as
\begin{equation}
   E_{\rm eff}
   =
   \sum_j n_j \, \epsilon_j
,
\end{equation}
where 
$ n_j $ represents the number of the nucleon occupied $j$th level. 
Only blocks containing the state with the lowest $E_{\rm eff}$ will be selected for diagonalization. 
For instance, let's consider a state $ \vert 0011 \rangle $ in the 2$^{\text{nd}}$ block (see Fig. \ref{fig2}(c)). 
Here, we have 
$ n_1 = 2 $, 
$ n_2 = 1 $, and 
$ n_3 = n_4 = 0 $ 
yielding 
$ E_{\rm eff} = 2 \times 1 + 1 \times 2 + 0 \times 3 + 0 \times 4 = 4 $ 
MeV. 
This value is the smallest value when compare with the other states. 
Thus, only the 2$^{\textrm {nd}}$ block need to diagonalize. 
It's worth noting that this method applies solely to odd configurations with the smallest seniority, 
i.e., at zero temperature. 
Because only the ground-state eigenvalue is needed at $T=0$. 
When extending the calculations to finite temperatures, 
all blocks must be diagonalized to ensure the statistical properties for the partition function of system.

\section{Description of the EP code}
\label{Use}

The EP code is written in Fortran programming language.
It is designed to be portable and flexible in use.
The computational core of the EP code is explained in the following subsection.

Before doing so, let us start with specifying the input of the EP code, namely 
$
{
   \tt 
   n 
}
= 
   \Omega
$,
$
{
   \tt 
   npar 
}
   = 
   N
$, 
$
   {
      \tt 
      g
   }
   (j,k)
=
   G_{jk}
$, 
and 
$
   {\tt e}
   (k)
=
   \epsilon_{k}
$, 
for $ j,k = \overline{1,\Omega} $.
In this context, $ \tt ir $ and $\tt ic $ are denoted as
the row and column indices, 
respectively,
of the matrix element 
$ H_{ \tt  ir, ic }$ 
in
$\hat{H}$,
where 
$
   1 
\le 
   \tt
   ir, ic
\le 
   \tt
   ljj
$, 
$
   \tt 
   ljj
=
   L
$,
and $L$ is defined in the section 
\ref{store}.

\subsection{Main calculations}

\label{main}

The main calculations of the EP code are described in the following steps:

\paragraph{Step 1. Generate all possible states}
The ${\bf gens}$ procedure ({\bf Alg.} \ref{alg-gens}) takes inputs $p = {\tt nps}$ to generate an array of all the possible POSs, stored as ${\tt njj}(m)$ for $m = \overline{ 1, L }$. Here, ${\tt nps} = \lfloor {\tt npar}/2 \rfloor - \lfloor {\tt ns}/2 \rfloor$, where ${\tt ns}$ (seniority) is 0 if ${\tt npar}$ is even, or 1 otherwise. 

\paragraph{Step 2.
Estimate the length of storage for essential arrays}
The $ {\bf npairing} $ function ({\bf Alg.} \ref{alg-npairing}) calculates the number of non-zero elements in the upper triangular part of $\hat{H}$. This step determines the storage size for the sparse matrix. The indices of the upper triangular $\hat{H}$ will be computed and stored in the MSR format within the next step.

\paragraph{Step 3.
Determine  
the upper triangular part of $\hat{H}$}
\label{H-jah-aah}
First, the ${\bf genp}$ procedure ({\bf Alg.} \ref{alg-genp}) generates the pair-scattering states for a given POS. Next, the ${\bf hashs}$ function ({\bf Alg.} \ref{alg-hashs}) retrieves the column indices of these states. Then, the ${\bf gind}$ subroutine ({\bf Alg.} \ref{alg-gind}) is used to compute the non-zero element values. Finally, the ${\bf deccon}$ procedure ({\bf Alg.} \ref{alg-dec2bin}) determines the main diagonal of $\hat{H}$.

\paragraph{Remark}
If the pairing strength ${\tt g}$ is a constant, the non-zero off-diagonal elements are all equal to $-{\tt g}$, so we only needs to store $L$ elements of the diagonal, significantly reducing RAM usage for large $\Omega$ and $N$.

\paragraph{Step 4.
Numerically solve the eigenpair problem for the first few eigenvalues}
As mentioned in the Section \ref{eigenpair}, for large $\hat{H}$, the ARPACK is used, requiring a matrix-vector multiplication procedure ({\bf Alg.} \ref{alg-matvec}), invoked via the DSAUPD subroutine. For small $\hat{H}$, the LAPACK subroutine with the eigenvector option is used directly. The \textbf{matvec} procedure ({\bf Alg.} \ref{alg-matvec}) handles the matrix-vector multiplication, designed for the general cases with $\tt g$ given as a matrix. If $\tt g$ is a scalar, the non-zero element values ${\tt aah}(k)$ in the algorithm are replaced by $-{\tt g}$, and ${\tt aah}$ is allocated with the length $L$ only.

\begin{center}
\begin{algorithm}[H]
\label{alg-matvec}
\caption{%
Procedure 
\textbf{matvec} 
performs 
the matrix-vector multiplication 
$\hat{H} \mathbf{x} $
and return the result to $\mathbf{y}$.
Here the $\hat{H}$ matrix is symmetric and given in the MSR format \cite{MSR}, 
where 
$\tt jah $
and 
$\tt aah$
are the arrays of index and non-zero elements, respectively. 
$L$ is the order of $\hat{H}$, 
and 
$ \mathbf{x} $ is the input vector.
}
\Input{ $\tt jah$, $\tt aah$,  $L$,  $ \mathbf{x} $ }
\Output{ $ \mathbf{y} $ }
\Local{ $ i, j, k $ } \; 
   \For{ 
      $ i \gets 1$   \KwTo  $ L $  
   }{
      $ 
         y (i) \gets  {\tt aah} (i) * x (i) 
      $ \; 
   }
   \For{ 
      $ i \gets 1$   \KwTo  $ L $  
   }{
      \For{ 
         $ k \gets {\tt jah} (i) $   \KwTo  $ {\tt jah} (i+1) - 1 $  
      }{
         $ j      \gets  {\tt jah} (k) $ \; 
         $ y(i)   \gets  y(i)  +  {\tt aah}(k) *x(j) $ \; 
         $ y(j)   \gets  y(j)  +  {\tt aah}(k) *x(i) $ \; 
      }
   }
\end{algorithm}
\end{center}
Finally, to numerically compute the occupation numbers in the Eq. \eqref{fEP0} for large systems, the Kahan summation algorithm is employed to reduce the effect of round-off errors; see, e.g., \cite{Higham}.

\subsection{Subroutine interfaces}
\label{subroutine}

The current EP code is designed to have only one main
subroutine with many interfaces, 
as shown in Listing \ref{lst1}.

%{%
%\lstset{basicstyle=\large} 
\lstset{basicstyle=\small,style=myFortranStyle} 
\begin{lstlisting}[caption={Generic user interface of EP},label={lst1}]
   subroutine ep_gstate ( & 
              nome, npar, e, [ng,] g, neiv, [tole,] prefix) 
\end{lstlisting}
%}%

We describe the arguments of the subroutine in Listing \ref{lst1} as follows:
\begin{itemize}

\item 
$ \texttt{nome}  $
is a 32-bit integer scalar that specifies the number of levels
$ \Omega $.

\item 
$ \texttt{npar} $
is a 32-bit integer scalar that defines the number of particles $N$.

\item 
$ \texttt{e} $ is a real array of dimension(\texttt{nome}). $\texttt{e}(j) = \epsilon_j$ is the single-particle energy, for  $ j = \overline{1, \Omega } $.  
\item
\texttt{ng} is an optional 32-bit integer scalar.

\item $\tt g$ is a real array of dimension($\tt ng,ng $) when $\tt ng$ is provided. In that case, $\tt g$ is either a constant ($\tt ng=1$) or a symmetric matrix ($\tt ng=nome$).  If $\tt ng$ is omitted, $\tt g $ must be a real scalar.

\item 
$ \texttt{neiv} $
is a 64-bit integer scalar that defines the number of eigenvalues we want to compute, $\texttt{neiv} \le L $.

\item 
$ \texttt{tole} $ 
is an optional real scalar that defines 
the numerical tolerance in the ARPACK.
If 
$ 
   \texttt{tole}  
$
is omitted, 
or provided as 
zero,
the EP code sets the tolerance to the machine epsilon. 

\item 
\texttt{prefix}
is a character array that defines the full path of output files.

\end{itemize}
Here 
$\tt g$,
$\tt e$,
and 
$\tt tole$
must be given in the same precision. 
Depending on whether these inputs are specified in 
%% single ($\tt real(4)$), 
double
(i.e., $\tt real(kind=8)$), 
or quadruple precision 
(i.e., $\tt real(kind=16)$), 
the program will automatically perform the real number calculations using the corresponding precision.
This feature is achieved via a generic interface block in Fortran programming
language after several subroutines, 
such as the 
DSAUPD, 
DSYEV, 
and the others involved in the 
ARPACK \cite{arpack},
LAPACK \cite{lapack},
and BLAS \cite{blas}, 
are modified slightly to support the different precision types.

\subsection{Configuration and Execution}
\label{exe}
After successfully compiling the EP code, an executable program is generated, which uses a configuration file to supply inputs for the \texttt{ep\_gstate} subroutine, as shown in Listing \ref{lst1}. A detailed  guide can be found in the {\tt README.txt} file.

\textit{Note:} The matrix ${\tt g}(:,:)$ (i.e., $G_{jj'}$ in Eq. (\ref{Vjj}))  is defined element by element in the format: {\tt i ~    j   ~ G(i,j)}, where ${\tt i,j =
\overline{1, nome}} $. Since the matrix is symmetric, only one triangular part (upper or lower) needs to be provided. Alternatively, the user
can input the pairing matrix elements $V_0(jj,j'j')$ using the same format as for
$G_{jj'}$. In this case, the code will automatically compute $G_{jj'}$ from $V_0$ using the Eq. (\ref{Vjj}).

Listing \ref{lst2} shows an example of a configuration file for the EP code execution.

\lstset{style=myTextStyle} 
\begin{lstlisting}[caption={Example of a configuration file for the EP code execution},label={lst2}]
nome = 4
npar = 4
g = 0.4
neiv = 1 
tole = 0 
prefix = test
e = 
1 
2 
3 
4 
#prep = True 
 
__EOF__ 
\end{lstlisting}

The EP code can be compiled and executed easily using a Linux {\tt Makefile} or the command lines. With the {\tt Makefile}, the {\tt make} command compiles the code, adjustable for the {\tt ifort} compiler by setting {\tt FC}. Alternatively, the command lines in Listing \ref{lst3} can be used step-by-step. The compiled executables ($\tt ep\_dp.exe$ for the double precision and $\tt ep\_qp.exe$ for the quadruple precision) are run with the configuration files (e.g., $ \tt test/ep\_conf1.txt $, $  \tt 
test/ep\_conf2.txt $), as shown in Listing \ref{lst4}. For Windows, the command-line compilation method requires replacing Unix slashes with backslashes in the file paths and omitting the Unix dot-slash for executable commands.

\lstset{style=myTextStyle} 
\begin{lstlisting}[caption={Manually compilation of the EP code in Linux},label={lst3}]
gfortran -O3 -o 1.o -c src/epcode_sub.f
gfortran -O3 -o 2.o -c src/epcode_dep_dp.f
gfortran -O3 -o 3.o -c src/epcode_dep_qp.f
gfortran -O3 -o 4.o -c src/epcode_mod.f
gfortran -O3 1.o 2.o 3.o 4.o -o ep_dp.exe src/epcode_main_dp.f90
gfortran -O3 1.o 2.o 3.o 4.o -o ep_qp.exe src/epcode_main_qp.f90
\end{lstlisting}

\lstset{style=myTextStyle} 
\begin{lstlisting}[caption={Execution of the EP code in Linux with configuration
files},label={lst4}]
./epcode_dp.exe test/ep_conf0.txt
or
./epcode_qp.exe ep_conf.txt
\end{lstlisting}

Finally, the eigenvalues and eigenvectors of the EP problem are stored in an external file with the suffix { \tt xxx$\_$result.txt}.

A pre-processing mode is also available and can be enabled using the {\tt prep = True} flag in the configuration file. In this mode, the program only calculates the matrix dimensions and estimates the required amount of RAM based on the specified values of $\Omega$ and N, then terminates. This allows users to determine whether their computer configuration meets the computational requirements.

Last but not least, a series of test cases is provided via configuration files named {\tt ep$\_$conf0.txt} to { \tt ep$\_$conf6.txt} in the {\tt test} folder. Each test is thoroughly described within its respective file and in the accompanying {\tt README.txt}. Notably, the {\tt ep$\_$conf5.txt} contains a test case for the realistic nucleus $^{114}$Sn, using the experimental single-particle energies and pairing matrix elements derived from the G-matrix calculation \cite{Zele}. Users can run all test cases by executing the command {\tt make test}.

\section{Results and discussion}
\label{Re}
\subsection{Computational configuration}
Our calculations are performed on a desktop computer  
with Intel(R) Core(TM) i7-9700F CPU @ 3.00GHz and 16GB RAM. 
The testing compilers are 
   the Intel Fortran compiler (Ifort; version 2021.10.0)
and 
   Gfortran compiler (version 12.2.0) 
on
   Debian GNU/Linux 12 (bookworm) system. 
Because of its popularity, the Gfortran compiler is primarily used, while the Ifort serves as a reference in this work. 
For simplicity, 
we employ the test configuration with a constant pairing strength 
$ \textbf{G} = G $, 
$ N $ 
nucleons and the single-particle energies $ \epsilon_j=j $ for $ j=1,...,\Omega $, 
where $ \Omega $ represents the number of deformed single-particle levels. 
To perform the calculations with realistic nuclear systems, 
users can extract $ \epsilon_j $ and $ \textbf{G} $ from the nuclear structure calculations such as shell model. 
In this work, the symmetrical configurations corresponding to $ \Omega = N $, 
which have the largest matrix size and 
the most non-zero elements compared to those with 
$ \Omega \neq N $, are used. 
Users have the flexibility to adjust the values of  
$ \Omega $ and $ N $ 
as needed when using the code. 
However, due to the 16GB RAM limitation in this work, 
the maximum value of $N$ is 26.

\subsection{Eigenvalue and eigenvector}

Once the pairing matrix is constructed, the \textbf{DSAUDP} subroutine of ARPACK \cite{arpack} is called to diagonalize this matrix with a specified accuracy level, which can be freely declared by users. For the matrices smaller than $ 100 \times 100 $, the \textbf{DSYEV} subroutine of the LAPACK \cite{lapack} is adopted to achieve better performance. The ground-state eigenvalues of specific configurations with $G=0.4$ obtained from the EP code are presented in Table \ref{table1} and \ref{table2}. 

Table \ref{table1} shows the ground-state eigenvalues of the problems with 
$
   G = 0.4
$, 
$
   (\Omega , N )
=
   ( 22, 10 ),
   ( 26, 10 ),
$
and 
$
   ( 30, 10 )
$
obtained from the EP code. As shown in this table, our results ($E_{\rm EP}$)
are comparable to those obtained from the PairDiag code ($E_{\rm PairDiag}$) and the polynomial algorithm solution of Richardson’s equations ($E_{\rm Richardson}$) \cite[Table 3]{PairDiag}. The results obtained from these calculations exhibit nearly identical values, with very small differences that are negligible. On the other hand, as we can see in Table \ref{table1}, $E_{\rm EP}$ fits $E_{\rm Richardson}$ 
more closely. This underscores the reliability of our code. The calculations for the (30,10) configuration exceed the 16GB RAM limit, requiring the use of swap space in Linux.
Consequently, this outcome is used solely for accuracy comparison and not for other evaluations.

\begin{table}[H]
\centering
\caption{%
\label{table1} 
Ground-state eigenvalues of the EP matrix with 
$
   G = 0.4
$
obtained from the EP code ($E_{\rm EP}$),
compared with those obtained from 
the PairDiag code ($E_{\rm PairDiag}$) 
and 
the polynomial algorithm solution of the Richardson's equations ($E_{\rm Richardson}$)
\cite[Table 3]{PairDiag}.
Herein,
$
   L
=
   \texttt{ljj} 
=
   C^{ \Omega - s }_{ N/2 } 
$
and 
$s=0$.
}
\begin{tabular}{|crccc|}
\hline 
   ($\Omega , N $)
%  & {\rm Dimension}   
   & $L$
   & $E_{\rm EP}$ 
   & $E_{\rm PairDiag}$ 
   & $E_{\rm Richardson}$ 
\\ 
\hline 
%
% DO NOT DELETE THIS if we still need it 
%  ~/tmp/epcode_res5/step7/exe.tst_DeformEP20.gfortran.dp_22
%
   $( 22, 10 )$
   &  $ 6.5 \times 10^5 $ 
   &  {103.0163819}
   &  {103.0163817}
   &  {103.0163818}
\\ 
\hline 
%
% DO NOT DELETE THIS if we still need it 
%  ~/tmp/epcode_res5/step7/exe.tst_DeformEP20.gfortran.dp_26
%
   $( 26, 10 )$
   &  $ 5.3 \times 10^6 $ 
   &  {102.2599361}
   &  {102.2599359}
   &  {102.2599361}
\\ 
\hline 
%
% DO NOT DELETE THIS if we still need it 
%  ~/tmp/epcode_res5/step7/exe.tst_DeformEP20.gfortran.dp_30
%
   $( 30, 10 )$
   &  $ 3.0 \times 10^7 $ 
   &  {101.5397386}
   &  {101.5397383}
   &  {101.5397386}
\\ 
\hline 
\end{tabular} 
\end{table}

Table \ref{table2} 
shows the ground-state eigenvalues of the EP matrix obtained from the EP and PairDiag codes in various cases, where  
$ G = 0.4 $ and  
$\Omega = N = \overline{10,26} $. 
   {$\tt EP_{qp}$}
   and 
   {$\tt EP_{dp}$}
   denote the EP eigenvalues performed with the quadruple ($\tt qp$) and double ($\tt dp$) precision, respectively. 
   The tolerance parameter is set as  
   ${\tt tole} = 0$.
   ${\tt RelErr}$ 
   stands for the relative error between the corresponding results and 
   {$\tt EP_{qp}$}.
Here, we show only the results obtained with the Gfortran compiler, 
as those obtained with the Ifort are identical. 
The table shows that the eigenvalues obtained from the PairDiag and EP codes match to 8 decimal digits. 
However, the relative error of the EP code remains stable, 
whereas that of the PairDiag code increases with increasing $\Omega$.

\begin{table}[H]
\centering
\caption{%
\label{table2} 
   Ground-state eigenvalues of the EP matrix obtained from 
   the EP and PairDiag codes as $ G = 0.4 $ and  $\Omega = N = \overline{10,26} $. 
   Here, 
   {$\tt EP_{qp}$}
   and 
   {$\tt EP_{dp}$}
   stand for the ground-state eigenvalues obtained from 
   the EP code with ${\tt tole} = 0$ in the quadruple and double precision, respectively.
   ${\tt RelErr}$ 
   stands for the relative error between the {$\tt EP_{dp}$}({\tt PairDiag}) and 
   {$\tt EP_{qp}$}.
   The calculations are performed with the Gfortran compiler. 
}
{%\footnotesize 
\scriptsize 
\begin{tabular}{|c|r|rr|rr|}
\hline 
   $ \Omega $ 
   & {$\tt EP_{qp}$}
   & {$\tt EP_{dp}$}
   & {\tt RelErr}
   & {\tt PairDiag} 
   & {\tt RelErr}
\\ 
\hline 
%
% DO NOT DELETE THIS if we still need it 
%  ./pl.get_result n1=10 n2=26 compiler=gfortran
%  res_EEv_gfort.tex 
%
%% N &               EP (qp) &          EP (dp) &           ReErrEiV &         PairDiag &           ReErrEiv
  10 &  27.10381384670832984 &   27.10381384671 & $1 \cdot 10^{-15}$ &   27.10381384671 & $3 \cdot 10^{-15}$ \\ 
  11 &  33.42689281267189311 &   33.42689281267 & $2 \cdot 10^{-16}$ &   33.42689281267 & $6 \cdot 10^{-16}$ \\ 
  12 &  38.42071511860793235 &   38.42071511861 & $8 \cdot 10^{-16}$ &   38.42071511861 & $6 \cdot 10^{-16}$ \\ 
  13 &  45.83415763874477369 &   45.83415763874 & $6 \cdot 10^{-16}$ &   45.83415763874 & $7 \cdot 10^{-16}$ \\ 
  14 &  51.70986480928340535 &   51.70986480928 & $1 \cdot 10^{-16}$ &   51.70986480928 & $4 \cdot 10^{-15}$ \\ 
  15 &  60.22118085129759263 &   60.22118085130 & $2 \cdot 10^{-15}$ &   60.22118085130 & $2 \cdot 10^{-15}$ \\ 
  16 &  66.97168008460883906 &   66.97168008461 & $6 \cdot 10^{-16}$ &   66.97168008461 & $2 \cdot 10^{-15}$ \\ 
  17 &  76.58827676767563443 &   76.58827676768 & $4 \cdot 10^{-16}$ &   76.58827676768 & $8 \cdot 10^{-15}$ \\ 
  18 &  84.20636388316873836 &   84.20636388317 & $7 \cdot 10^{-16}$ &   84.20636388317 & $2 \cdot 10^{-14}$ \\ 
  19 &  94.93541252935004967 &   94.93541252935 & $5 \cdot 10^{-16}$ &   94.93541252935 & $1 \cdot 10^{-14}$ \\ 
  20 & 103.41400463120281208 &  103.41400463120 & $2 \cdot 10^{-15}$ &  103.41400463120 & $4 \cdot 10^{-14}$ \\ 
  21 & 115.26231348985022156 &  115.26231348985 & $2 \cdot 10^{-15}$ &  115.26231348985 & $1 \cdot 10^{-14}$ \\ 
  22 & 124.59463121243389870 &  124.59463121244 & $9 \cdot 10^{-15}$ &  124.59463121239 & $3 \cdot 10^{-13}$ \\ 
  23 & 137.56853401378627719 &  137.56853401379 & $2 \cdot 10^{-15}$ &  137.56853401371 & $5 \cdot 10^{-13}$ \\ 
  24 & 147.74824343922811259 &  147.74824343923 & $8 \cdot 10^{-16}$ &  147.74824343881 & $3 \cdot 10^{-12}$ \\ 
  25 & 161.85351015173818270 &  161.85351015174 & $1 \cdot 10^{-15}$ &  161.85351015102 & $4 \cdot 10^{-12}$ \\ 
  26 & 172.87482861518104560 &  172.87482861518 & $6 \cdot 10^{-15}$ &  172.87482861184 & $2 \cdot 10^{-11}$ \\ 
\hline 
\end{tabular} 
}%
\end{table}

\begin{figure}[!h]
\begin{center}
\includegraphics[width=1\textwidth]{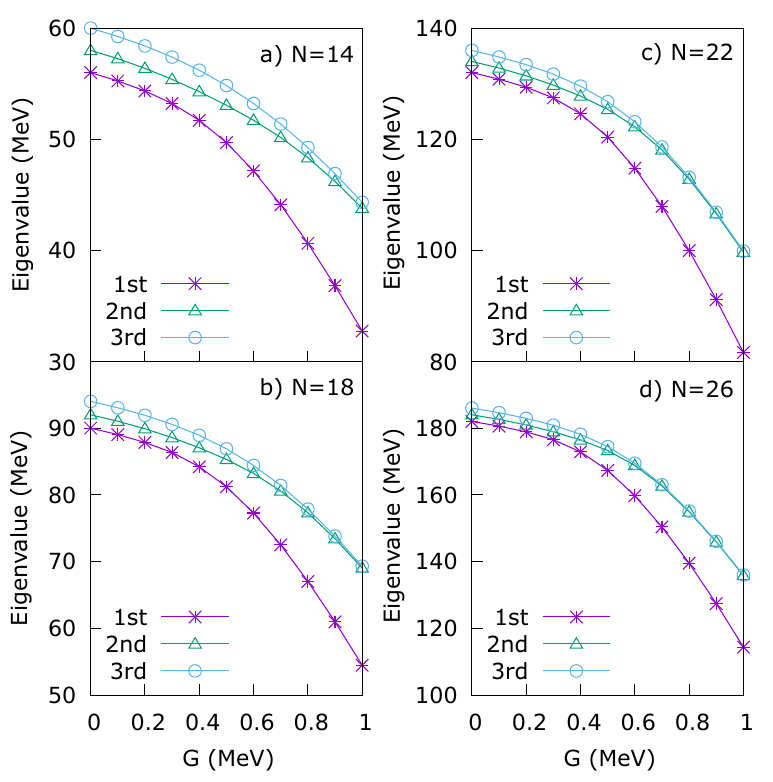}
\end{center}
\caption{\label{fig3} %
   The first three eigenvalues of the configurations 
   $ \Omega = N = 14, \, 18,\, 22, \, 26 $ obtained from the EP code for various $G$ values ranging from 0 to 1.
}
\end{figure}

\begin{figure}[!h]
\begin{center}
\includegraphics[width=1\textwidth]{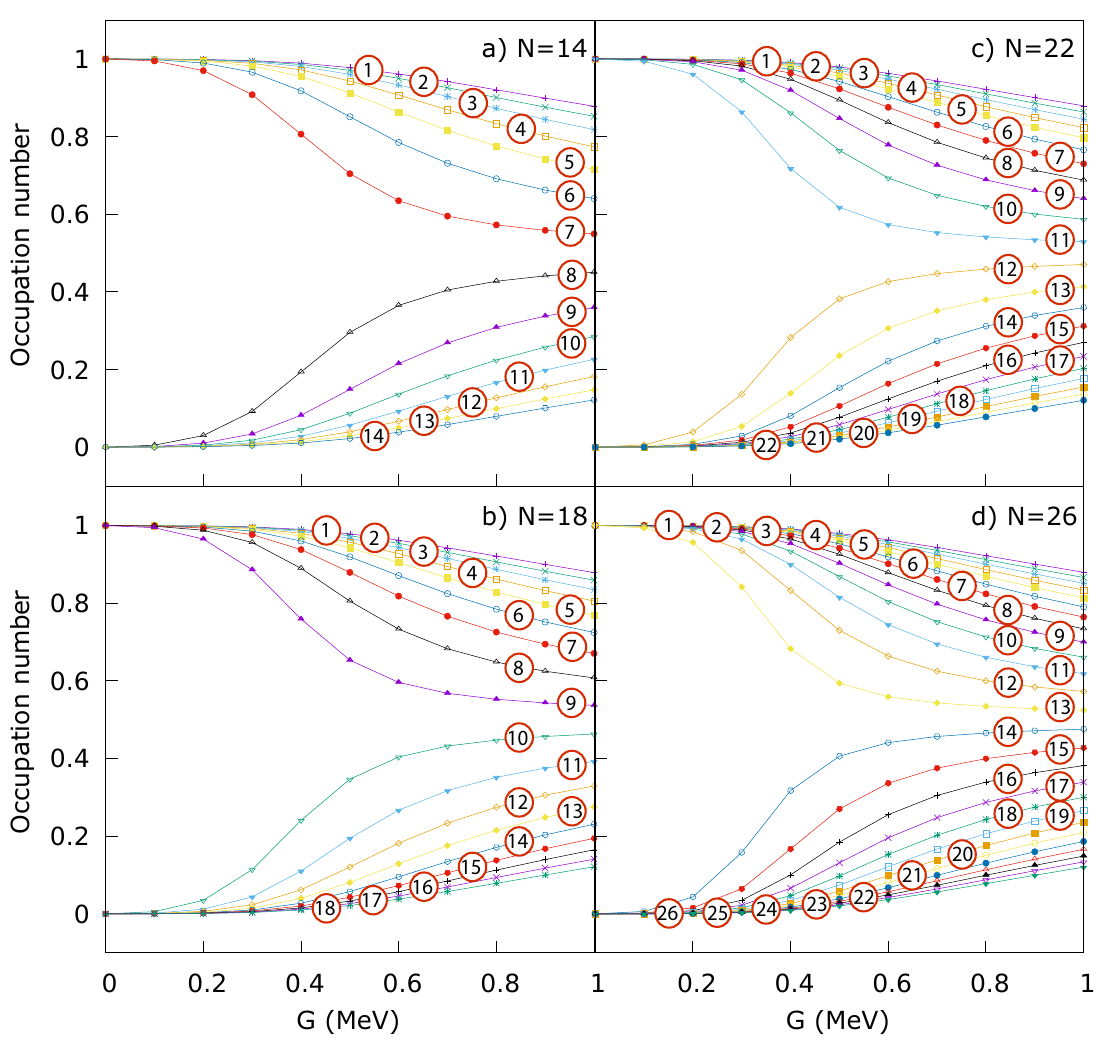}
\end{center}
\caption{%
\label{fig4} 
Occupation numpers of the ground-state of the configurations 
$ \Omega = N = 14, \, 18, \, 22, \, 26 $ 
versus pairing strength $G$.
The number in the circle indicates the number of level, i.e., 
$\Omega_j$, for $j = \overline{1,N} $.
}
\end{figure}

\begin{figure}[!h]
\begin{center}
\includegraphics[width=1\textwidth]{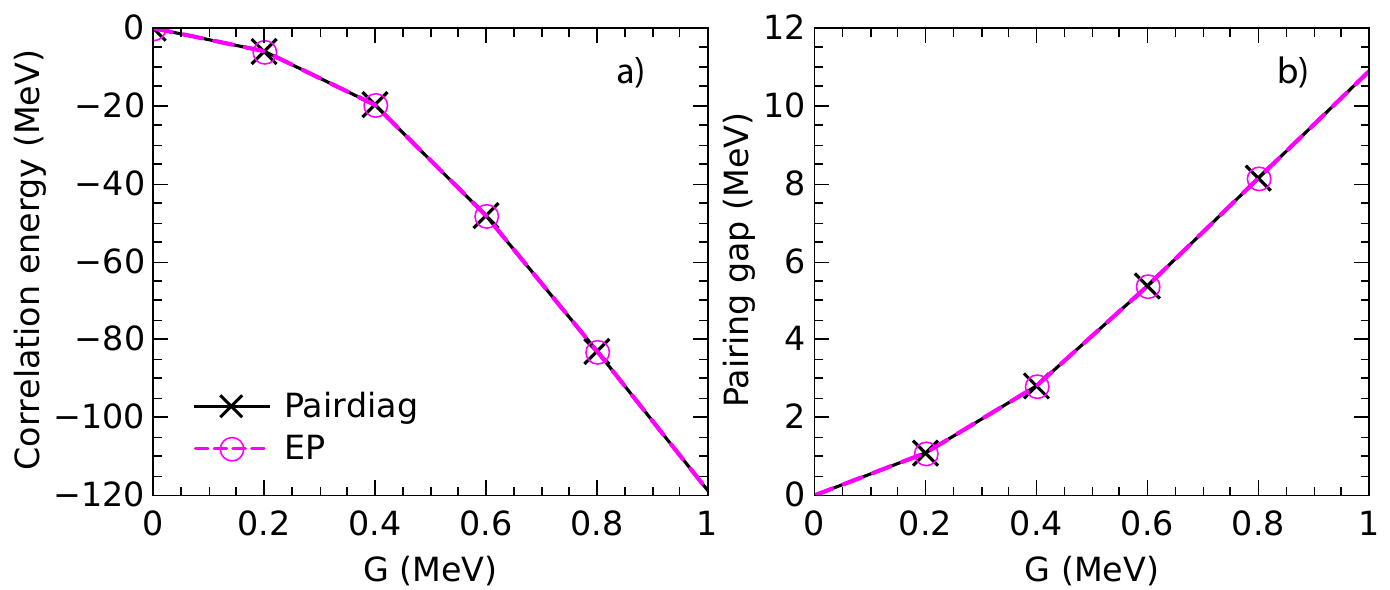}
\end{center}
\caption{%
\label{fig5} %
   The pairing correlation energies (a) and gap (b) of the system 
   obtained from the EP and PairDiag codes for the case where 
   $\Omega = N = 26$, and $G$ are varied from 0 to 1 MeV.
}
\end{figure}

In addition to the ground-state eigenvalues, our code readily computes all the eigenvalues of the pairing matrix by specifying the number of eigenvalues to be extracted, denoted as \textbf{neiv}. Fig. \ref{fig3} shows the first three eigenvalues of the pairing matrix (\textbf{neiv}=3) computed for various $G$ values ranging from 0 to 1 with $ \Omega = N = 14, \, 18,\, 22, \, 26 $. The eigenvectors corresponding to the ground-state eigenvalues of the configurations calculated in Fig. \ref{fig3} are presented in Fig. \ref{fig4}. In each eigenvector, the $j$th component represents the value of the single-particle occupation number on the level $j$. Fig. \ref{fig4} shows that as the pairing strength in the system increases, the nucleon pairs scatter more strongly to the single-particle levels above the Fermi level. The pairing correlation energies are also computed for comparison with the PairDiag code within the configuration $ \Omega = N = 26 $ and $ G = 0,...,1 $. The pairing correlation energies and gaps depicted in Fig. \ref{fig5} confirm the equivalence of numerical calculations between the two codes.

%%%%%%%%%%%%%%%%%%%%%%%%%%%%%%%%%%%%%%%%%%%%%%%%%%%%%%%%%%%%%%%%%%%%%%%
%%  DO NOT DELETE THIS if we still need it 
%#  To prepare for data from calculations 
%#  (e.g., with many values of G in [0,1] and N = Omega = 14,18,22,26)
%#  plot (by gnuplot, python), or extract everything etc. 
%#  0. run
%#    ./pl.get_result n1=10 n2=26 compiler=gfortran
%#    ./pl.get_result n1=10 n2=26 compiler=ifort 
%%  1. run  
%#    for n in 14 18 22 26 ; do perl pl.work $n  ;  done 
%#    bash sh.work_eival
%#    for f in *.gnu ; do gnuplot $f ; done
%#  2. use new scripts and data mentioned in figures/tables.  
%%%%%%%%%%%%%%%%%%%%%%%%%%%%%%%%%%%%%%%%%%%%%%%%%%%%%%%%%%%%%%%%%%%%%%%

%%%%%%%%%%%%%%%%%%%%%% DO NOT DELETE THIS if we still need it 
%
%  Figure 3:
%
%  data: 
%     res_occ-vs-g_N14_1  
%     res_occ-vs-g_N18_1    
%     res_occ-vs-g_N22_1    
%     res_occ-vs-g_N26_1  
%  plot: 
%     bash res_occ-vs-g.sh
%
%
%%%%%%%%%%%%%%%%%%%%%%%

%%%%%%%%%%%%%%%%%%%%%% DO NOT DELETE THIS if we still need it 
%
%  Figure 4:
%
%  data: 
%     res_eival-vs-g_N14  
%     res_eival-vs-g_N18  
%     res_eival-vs-g_N22  
%     res_eival-vs-g_N26
%  plot: 
%     bash res_eival-vs-g.sh
%
%
%%%%%%%%%%%%%%%%%%%%%%%
%%%%%%%%%%%%%%%%%%%%%% DO NOT DELETE THIS if we still need it 
%
%  Figure 5:
%
%  data: 
%     res_EPvsPD_N24_manyG.txt
%  plot: 
%     gnuplot res_EPvsPD_N24_manyG.gnu 
%
%
%%%%%%%%%%%%%%%%%%%%%%%

\begin{table}[H]
\centering
\caption{%
\label{tab-tole} 
The error estimate
$
   \max _{ 10 \le \Omega \le 26 }
   \| \hat{H}_{\Omega } \mathbf{v}_{\Omega }  - \lambda _{\Omega } \mathbf{v}_{\Omega } \| 
   /
   | \lambda_{\Omega } | 
$ 
obtained from the EP code executed with 
$ G = 0.4 $
and  
$\Omega = N $.
Here, 
for every $ \Omega = \overline{10, 26} $, 
the symbols 
$ 
   \hat{H}_{\Omega } 
$, 
$ 
   \mathbf{v}_{\Omega }
$,
and 
$ 
   \lambda _{\Omega } 
$
stand for the EP matrix, 
the ground-state eigenvalue, 
and  
the corresponding eigenvector,
respectively.
}
{%\footnotesize 
\scriptsize 
\begin{tabular}{|r|r|r|r|}
\hline 
   Compiler
   & { {$\tt EP_{qp}$} with $\tt tole = 0$} 
   & { {$\tt EP_{dp}$} with $\tt tole = 0$} 
   & { {$\tt EP_{dp}$} with $\tt tole = 10^{-8}$}
\\ 
\hline 
\hline 
   Gfortran
   &
   4.232$\cdot 10^{-32} $
   &
   5.301$\cdot 10^{-14} $
   &
   9.303$\cdot 10^{-9} $
\\
\hline 
   Ifort
   &
   3.265$\cdot 10^{-32} $
   & 
   4.059$\cdot 10^{-14} $
   & 
   9.303$\cdot 10^{-9} $
\\
\hline 
\end{tabular} 
}%
\end{table}

Table \ref{tab-tole} 
provides the error estimate
$
   \max _{ 10 \le \Omega \le 26 }
   \| \hat{H}_{\Omega } \mathbf{v}_{\Omega }  - \lambda _{\Omega } \mathbf{v}_{\Omega } \| 
   /
   | \lambda_{\Omega } | 
$  
obtained from the EP code with 
$ G = 0.4 $
and  
$\Omega = N $.
Here, 
the symbols 
$ 
   \hat{H}_{\Omega } 
$, 
$ 
   \mathbf{v}_{\Omega }
$,
and 
$ 
   \lambda _{\Omega } 
$
stand for the EP matrix, the ground-state eigenvalue, and the corresponding eigenvector, respectively. As we can see in this table, 
the errors of our eigenpair solver are reasonable, as expected, 
which implies that our numerical results are reliable.

\begin{table}[H]
\centering
\caption{%
\label{table4} 
   Error estimates for the ground-state eigenvalue
   and corresponding occupation numbers
   obtained 
   from the EP and
   PairDiag codes in three cases:
   (A)
   {$\tt EP_{dp}$}
   with $\tt tole = 0$, 
   (B) 
   {$\tt EP_{dp}$}
   with $\tt tole = 10^{-8}$, 
   and 
   (C)
   PairDiag with its default settings.
%% , i.e.,  
%%    ${\rm Lanc\_Limit} = 50$
%%    and 
%%    ${\rm Lanc\_Error} = 10^{-5}$.
   The calculated configurations are the same with
   Table \ref{table2}.
   Where, 
   ${\rm Err}_{\tt Eiv}$ and ${\rm Err}_{\tt Occ}$ 
   stand for the relative error of the 
   ground-state eigenvalue
   and 
   the maximum relative error of the corresponding occupation numbers, respectively.
}
{%\footnotesize 
\scriptsize 
\begin{tabular}{|c|rr|rr|rr|}
\hline 
   &  
      \multicolumn{2}{c|}{(A)
         {$\tt EP_{dp}$}
         with $\tt tole = 0$
      }
   &  
      \multicolumn{2}{c|}{(B)  
         {$\tt EP_{dp}$}
         with $\tt tole = 10^{-8}$ 
      }
   &  
      \multicolumn{2}{c|}{(C)
         PairDiag
      }
\\
\cline{2-7}
   $ \Omega $ 
   &  %(A) ~ 
      ${\rm Err}_{\tt Eiv}$
   &  ${\rm Err}_{\tt Occ}$
   &  %(B) ~
      ${\rm Err}_{\tt Eiv}$
   &  ${\rm Err}_{\tt Occ}$
   &  %(C) ~
      ${\rm Err}_{\tt Eiv}$
   &  ${\rm Err}_{\tt Occ}$
\\ 
\hline 
% DO NOT DELETE THIS if we still need it 
%  ./pl.get_result n1=10 n2=26 compiler=gfortran
%  res_EEO_gfort.tex 
%
%% N &            ErEiV 1 &            ErOcc 1 &            ErEiv 6 &            ErOcc 6 &            ErEiv 2 &            ErOcc 2 
  10 & $1 \cdot 10^{-15}$ & $1 \cdot 10^{-14}$ & $1 \cdot 10^{-15}$ & $5 \cdot 10^{-08}$ & $3 \cdot 10^{-15}$ & $2 \cdot 10^{-14}$ \\ 
  11 & $2 \cdot 10^{-16}$ & $3 \cdot 10^{-15}$ & $2 \cdot 10^{-16}$ & $4 \cdot 10^{-09}$ & $6 \cdot 10^{-16}$ & $2 \cdot 10^{-14}$ \\ 
  12 & $8 \cdot 10^{-16}$ & $1 \cdot 10^{-14}$ & $8 \cdot 10^{-16}$ & $5 \cdot 10^{-09}$ & $6 \cdot 10^{-16}$ & $4 \cdot 10^{-14}$ \\ 
  13 & $6 \cdot 10^{-16}$ & $6 \cdot 10^{-15}$ & $6 \cdot 10^{-16}$ & $7 \cdot 10^{-08}$ & $7 \cdot 10^{-16}$ & $4 \cdot 10^{-14}$ \\ 
  14 & $1 \cdot 10^{-16}$ & $1 \cdot 10^{-14}$ & $1 \cdot 10^{-16}$ & $2 \cdot 10^{-08}$ & $4 \cdot 10^{-15}$ & $9 \cdot 10^{-12}$ \\ 
  15 & $2 \cdot 10^{-15}$ & $3 \cdot 10^{-15}$ & $2 \cdot 10^{-15}$ & $1 \cdot 10^{-07}$ & $2 \cdot 10^{-15}$ & $9 \cdot 10^{-13}$ \\ 
  16 & $6 \cdot 10^{-16}$ & $4 \cdot 10^{-14}$ & $9 \cdot 10^{-16}$ & $2 \cdot 10^{-07}$ & $2 \cdot 10^{-15}$ & $5 \cdot 10^{-10}$ \\ 
  17 & $4 \cdot 10^{-16}$ & $3 \cdot 10^{-14}$ & $9 \cdot 10^{-16}$ & $8 \cdot 10^{-08}$ & $8 \cdot 10^{-15}$ & $4 \cdot 10^{-11}$ \\ 
  18 & $7 \cdot 10^{-16}$ & $3 \cdot 10^{-14}$ & $7 \cdot 10^{-16}$ & $9 \cdot 10^{-08}$ & $2 \cdot 10^{-14}$ & $7 \cdot 10^{-09}$ \\ 
  19 & $5 \cdot 10^{-16}$ & $4 \cdot 10^{-14}$ & $5 \cdot 10^{-16}$ & $8 \cdot 10^{-08}$ & $1 \cdot 10^{-14}$ & $1 \cdot 10^{-09}$ \\ 
  20 & $2 \cdot 10^{-15}$ & $3 \cdot 10^{-14}$ & $2 \cdot 10^{-15}$ & $9 \cdot 10^{-08}$ & $4 \cdot 10^{-14}$ & $7 \cdot 10^{-08}$ \\ 
  21 & $2 \cdot 10^{-15}$ & $4 \cdot 10^{-14}$ & $2 \cdot 10^{-15}$ & $1 \cdot 10^{-07}$ & $1 \cdot 10^{-14}$ & $1 \cdot 10^{-08}$ \\ 
  22 & $9 \cdot 10^{-15}$ & $4 \cdot 10^{-14}$ & $8 \cdot 10^{-16}$ & $2 \cdot 10^{-08}$ & $3 \cdot 10^{-13}$ & $3 \cdot 10^{-07}$ \\ 
  23 & $2 \cdot 10^{-15}$ & $1 \cdot 10^{-13}$ & $5 \cdot 10^{-15}$ & $1 \cdot 10^{-07}$ & $5 \cdot 10^{-13}$ & $6 \cdot 10^{-08}$ \\ 
  24 & $8 \cdot 10^{-16}$ & $3 \cdot 10^{-13}$ & $6 \cdot 10^{-15}$ & $5 \cdot 10^{-08}$ & $3 \cdot 10^{-12}$ & $6 \cdot 10^{-07}$ \\ 
  25 & $1 \cdot 10^{-15}$ & $2 \cdot 10^{-13}$ & $1 \cdot 10^{-15}$ & $2 \cdot 10^{-07}$ & $4 \cdot 10^{-12}$ & $1 \cdot 10^{-07}$ \\ 
  26 & $6 \cdot 10^{-15}$ & $3 \cdot 10^{-13}$ & $6 \cdot 10^{-15}$ & $3 \cdot 10^{-08}$ & $2 \cdot 10^{-11}$ & $3 \cdot 10^{-06}$ \\ 
\hline 
\end{tabular} 
}
\end{table}

Table \ref{table4}
shows 
   the error estimates for the ground-state eigenvalue
   and corresponding occupation numbers
   obtained 
   from the EP and
   PairDiag codes in three cases:
   (A)
   {$\tt EP_{dp}$}
   with $\tt tole = 0$, 
   (B)
   {$\tt EP_{dp}$}
   with $\tt tole = 10^{-8}$, 
   and 
   (C)
   PairDiag with its default settings.
%% , i.e.,  
%%    ${\rm Lanc\_Limit} = 50$
%%    and 
%%    ${\rm Lanc\_Error} = 10^{-5}$.
   The calculations of these cases are performed with 
   the Gfortran compiler.
Here, ${\rm Err}_{\tt Eiv}$ and ${\rm Err}_{\tt Occ}$ denote the relative error of the ground-state eigenvalue and the maximum relative error of the corresponding occupation numbers, respectively. These errors are computed by comparing the results of (A), (B), and (C) with the reference result obtained from the EP code executed in quadruple precision with the setting $\tt tole = 0$, which is considered the “exact” result due to its higher numerical precision. Table \ref{table4} shows the decrease in accuracy of the results obtained from the PairDiag code, whereas those obtained from the EP code remain stable with increasing $ \Omega $. This stability comes from the efficient use of the ARPACK package as well as the Kahan error compensation algorithm \cite{Higham} when calculating the occupation number from Eq. \eqref{fEP0} for large-scale configurations. A similar result is also observed when calculating with the Ifort compiler and is presented in Table \ref{table4-ifort}.

\begin{table}[H]
\centering
\caption{%
\label{table4-ifort} 
   Similarly to Table \ref{table4}, 
   the calculations in this table are performed with the Ifort compiler.
   Herein, "N.A." stands for "Not available".
}
{\footnotesize 
\begin{tabular}{|c|rr|rr|rr|}
\hline 
   $ \Omega $ 
   &  (A) ~ 
      ${\rm Eiv}$
   &  ${\rm Occ}$
   &  (B) ~
      ${\rm Eiv}$
   &  ${\rm Occ}$
   &  (C) ~
      ${\rm Eiv}$
   &  ${\rm Occ}$
\\ 
\hline 
% DO NOT DELETE THIS if we still need it 
%  ./pl.get_result n1=10 n2=26 compiler=ifort 
%  res_EEO_ifort.tex 
%
%% N &            ErEiV 1 &            ErOcc 1 &            ErEiv 6 &            ErOcc 6 &            ErEiv 2 &            ErOcc 2 
  10 & $3 \cdot 10^{-15}$ & $4 \cdot 10^{-14}$ & $7 \cdot 10^{-16}$ & $5 \cdot 10^{-08}$ & $1 \cdot 10^{-15}$ & $3 \cdot 10^{-15}$ \\ 
  11 & $2 \cdot 10^{-16}$ & $2 \cdot 10^{-14}$ & $5 \cdot 10^{-16}$ & $4 \cdot 10^{-09}$ & N.A.                &               N.A.  \\ 
  12 & $8 \cdot 10^{-16}$ & $6 \cdot 10^{-15}$ & $3 \cdot 10^{-16}$ & $5 \cdot 10^{-09}$ & $2 \cdot 10^{-15}$ & $1 \cdot 10^{-14}$ \\ 
  13 & $6 \cdot 10^{-16}$ & $2 \cdot 10^{-14}$ & $8 \cdot 10^{-16}$ & $7 \cdot 10^{-08}$ & N.A.                &               N.A.  \\ 
  14 & $1 \cdot 10^{-16}$ & $2 \cdot 10^{-14}$ & $3 \cdot 10^{-16}$ & $2 \cdot 10^{-08}$ & $1 \cdot 10^{-16}$ & $9 \cdot 10^{-12}$ \\ 
  15 & $1 \cdot 10^{-16}$ & $6 \cdot 10^{-15}$ & $1 \cdot 10^{-16}$ & $1 \cdot 10^{-07}$ & N.A.                &               N.A.  \\ 
  16 & $4 \cdot 10^{-15}$ & $1 \cdot 10^{-14}$ & $2 \cdot 10^{-15}$ & $2 \cdot 10^{-07}$ & $2 \cdot 10^{-15}$ & $5 \cdot 10^{-10}$ \\ 
  17 & $4 \cdot 10^{-16}$ & $2 \cdot 10^{-14}$ & $2 \cdot 10^{-16}$ & $8 \cdot 10^{-08}$ & $3 \cdot 10^{-15}$ & $4 \cdot 10^{-11}$ \\ 
  18 & $5 \cdot 10^{-16}$ & $2 \cdot 10^{-14}$ & $1 \cdot 10^{-15}$ & $9 \cdot 10^{-08}$ & $4 \cdot 10^{-15}$ & $7 \cdot 10^{-09}$ \\ 
  19 & $3 \cdot 10^{-15}$ & $1 \cdot 10^{-13}$ & $2 \cdot 10^{-16}$ & $8 \cdot 10^{-08}$ & $3 \cdot 10^{-15}$ & $1 \cdot 10^{-09}$ \\ 
  20 & $2 \cdot 10^{-15}$ & $4 \cdot 10^{-14}$ & $2 \cdot 10^{-15}$ & $9 \cdot 10^{-08}$ & $8 \cdot 10^{-15}$ & $7 \cdot 10^{-08}$ \\ 
  21 & $2 \cdot 10^{-15}$ & $3 \cdot 10^{-14}$ & $7 \cdot 10^{-16}$ & $1 \cdot 10^{-07}$ & $2 \cdot 10^{-14}$ & $1 \cdot 10^{-08}$ \\ 
  22 & $8 \cdot 10^{-16}$ & $1 \cdot 10^{-13}$ & $1 \cdot 10^{-17}$ & $2 \cdot 10^{-08}$ & $9 \cdot 10^{-15}$ & $3 \cdot 10^{-07}$ \\ 
  23 & $5 \cdot 10^{-15}$ & $2 \cdot 10^{-13}$ & $2 \cdot 10^{-15}$ & $1 \cdot 10^{-07}$ & $5 \cdot 10^{-14}$ & $6 \cdot 10^{-08}$ \\ 
  24 & $6 \cdot 10^{-15}$ & $1 \cdot 10^{-12}$ & $5 \cdot 10^{-15}$ & $5 \cdot 10^{-08}$ & $1 \cdot 10^{-13}$ & $6 \cdot 10^{-07}$ \\ 
  25 & $1 \cdot 10^{-15}$ & $3 \cdot 10^{-12}$ & $5 \cdot 10^{-16}$ & $2 \cdot 10^{-07}$ & $2 \cdot 10^{-13}$ & $1 \cdot 10^{-07}$ \\ 
  26 & $3 \cdot 10^{-16}$ & $8 \cdot 10^{-12}$ & $3 \cdot 10^{-15}$ & $3 \cdot 10^{-08}$ & $2 \cdot 10^{-12}$ & $3 \cdot 10^{-06}$ \\ 
\hline 
\end{tabular} 
}
\end{table}

\subsection{Evaluate running time}
\begin{figure}[H]
\begin{center}
\includegraphics[width=1\textwidth]{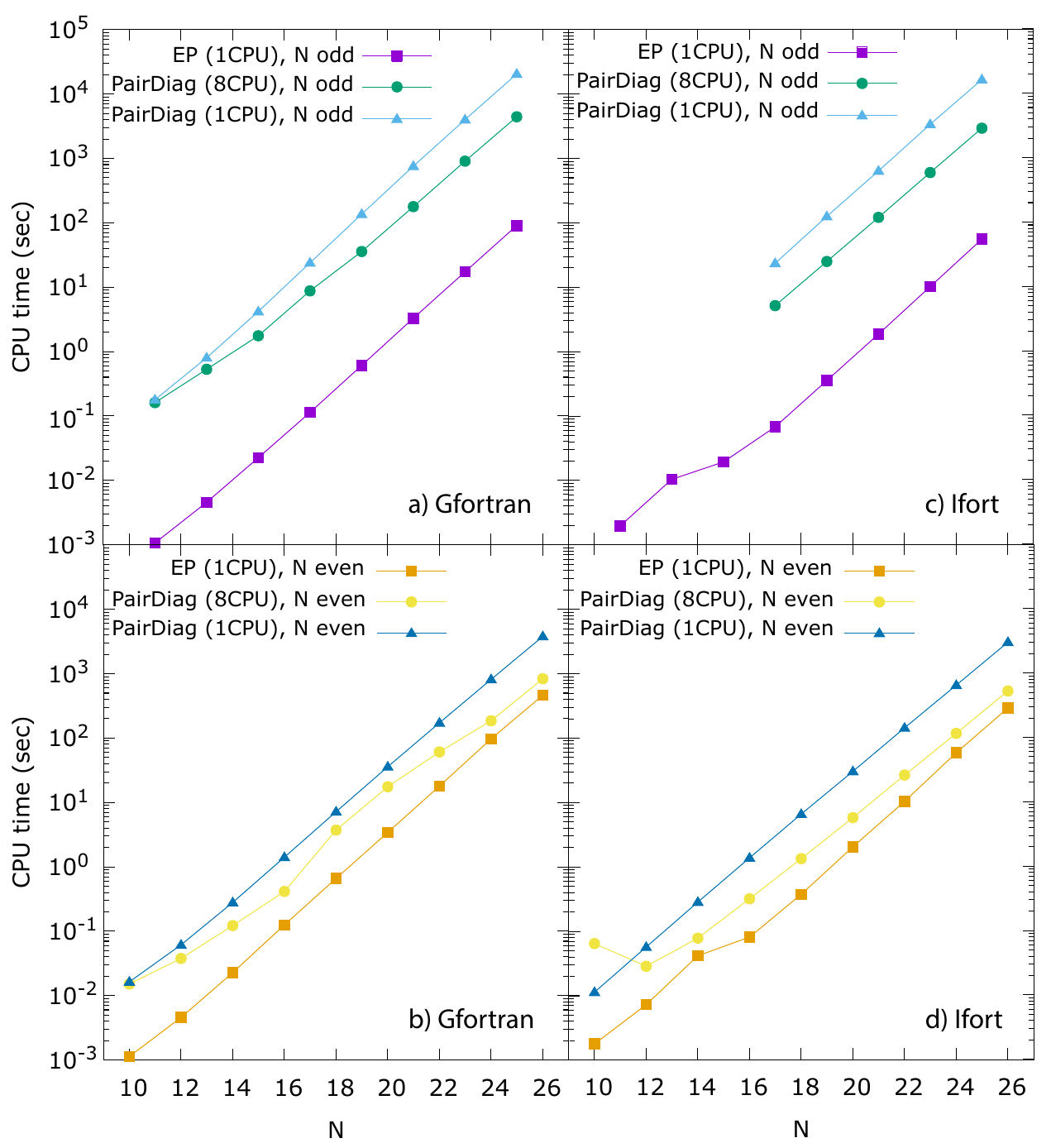}
\end{center}
\caption{%
   CPU time 
   (log10-scaled; in seconds) 
   consumed by the EP and PairDiag codes 
   for various configurations, i.e., $ N = \Omega $ and $N = \overline{10,26}$,  
   with different compilers, i.e., 
   the Gfortran ((a) and (b)),
   and the Ifort ((c) and (d)).
}
\label{fig6}
\end{figure}

Constructing the pairing matrix with reasonable breakpoints for loops and
working only on the upper half of the matrix yields substantial savings in
computational resources, thus reducing the computational time significantly. The efficiency of using the ARPACK package to diagonalize the sparse matrices with very large dimensions also contributes significantly to shortening calculation time.
The calculation speed, expressed in terms of the CPU time for various configurations
with \textbf{neiv} = 1, is depicted in Fig. \ref{fig6}. This figure shows the
CPU time (log10-scaled; in seconds) consumed by the EP and PairDiag codes for
various configurations $ N = \Omega = \overline{10,26}$, with the two different
compilers, i.e., Gfortran in Fig. \ref{fig6}(a) and Ifort in Fig. \ref{fig6}(b). In
Fig. \ref{fig6}(b), the PairDiag code failed with 
$N=11, \, 13, \, 15$. For the N-even configurations, as shown in both Fig. \ref{fig6}(a) and \ref{fig6}(b), the calculations with the EP code exhibit significantly faster performance compared to the PairDiag code when running on a single core. Remarkably, even the single-core calculation speed of the EP code surpasses that of the PairDiag code when running with 8 cores. This speed discrepancy becomes more pronounced as N increases. Specifically, the EP code running on a single core is nearly 8 times faster than the PairDiag code running on a single core, and 3 times faster than the PairDiag code running with 8 cores. For the N-odd configurations (Fig. \ref{fig6}(a)), this difference in the CPU time increases significantly. This significant difference stems from our technique of identifying and diagonalizing a single block, unlike the PairDiag code, which processes all blocks. For instance, at $ N=25 $, the CPU time of the EP code running on a single core is nearly 230 times shorter than that of the PairDiag code running on a single core and 50 times shorter than that of the PairDiag code running on 8 cores within the Gfortran compiler. These results are nearly 300 and 50 times when executed with the Ifort compiler.

%%%%%%%%%%%%%%%%%%%%%% DO NOT DELETE THIS if we still need it 
%
%  Figure 6:
%
%  data:
%     res_CPU_gfort.txt
%     res_CPU_ifort.txt
%  plot:
%     gnuplot res_CPU_gfort.gnu
%     gnuplot res_CPU_ifort.gnu
%
%
%
%%%%%%%%%%%%%%%%%%%%%%

\begin{figure}[H]
\begin{center}
\includegraphics[width=1\textwidth]{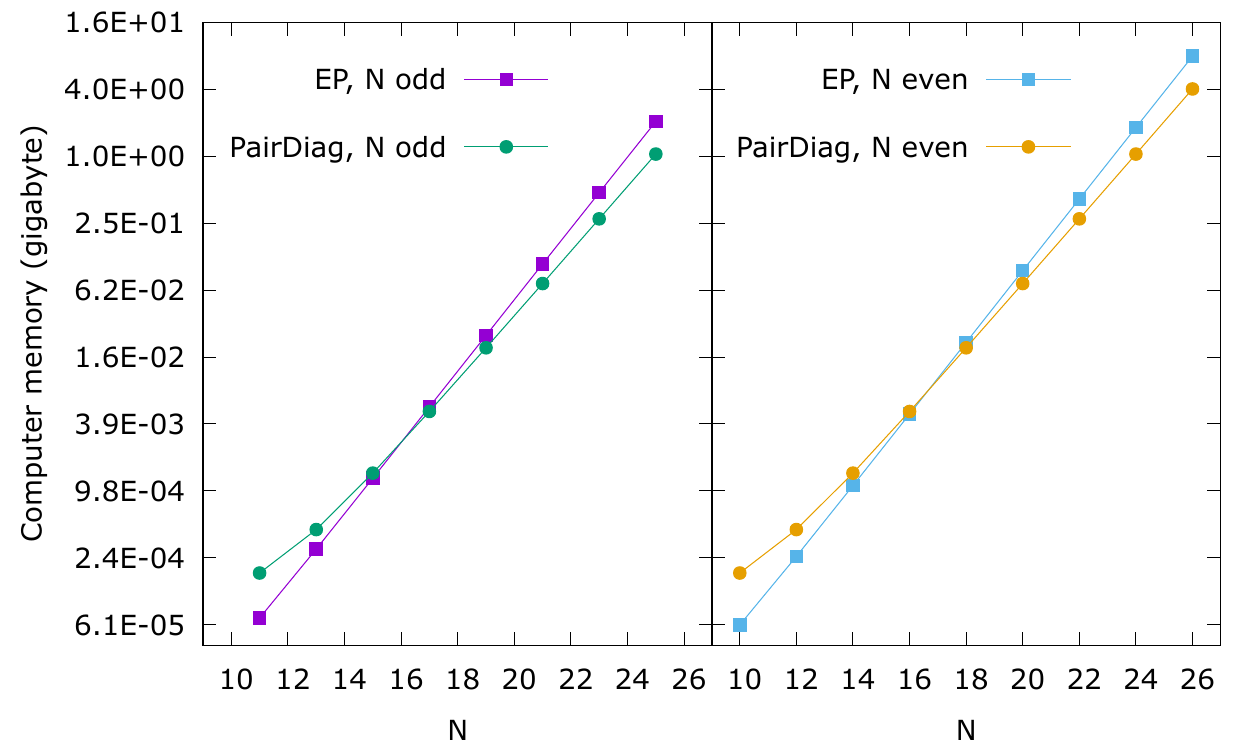}
\end{center}
\caption{%
   Computer memory estimation (log2-scaled; in gigabytes) obtained from the EP
and PairDiag codes using Gfortran compiler with the configurations 
$ N = \Omega = \overline{10, 26} $, 
where the case with $N$ odd (even) is shown in the left (right) subfigure.
}
\label{fig7}
\end{figure}

Fig. \ref{fig7} shows the computer memory estimation (log2-scaled; in gigabytes) obtained from the EP and PairDiag codes in various configurations. In general, the PairDiag code, which does not save the matrix elements on RAM (on the fly approach \cite{PairDiag}), conserves more memory than the EP code.

%%%%%%%%%%%%%%%%%%%%%% DO NOT DELETE THIS if we still need it 
%
%  Figure: 7
%
%  data:
%     res_RAM_gfort.txt
%  plot:
%     gnuplot res_RAM_gfort.gnu
%
%
%
%%%%%%%%%%%%%%%%%%%%%%

%%%%%%%%%%%%%%%%%%%%%% DO NOT DELETE THIS if we still need it 
%
%  Figure 8:
%
%  data:
%     res_CPU_gfort.txt
%     res_CPU_ifort.txt
%  plot:
%     gnuplot res_EPt_gfort.gnu
%     gnuplot res_EPt_ifort.gnu
%
%
\begin{figure}[!h]
\begin{center}
\includegraphics[width=1\textwidth]{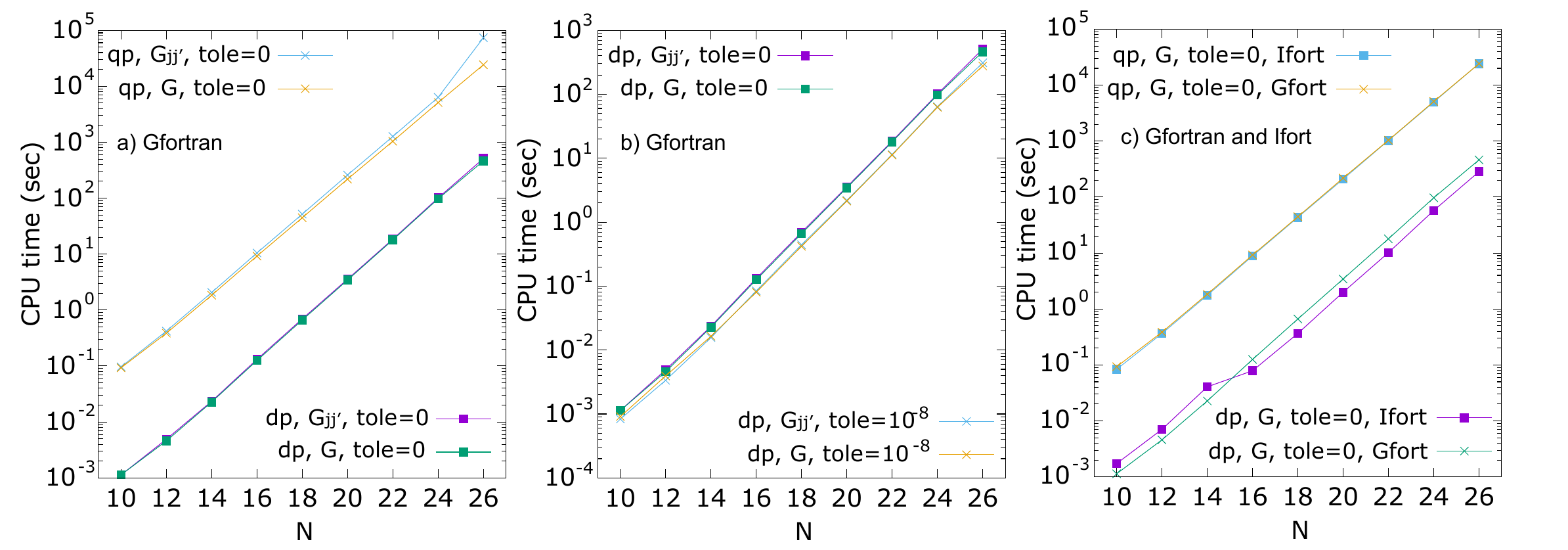}
\end{center}
\caption{%
\label{fig8} %
   CPU time consumed by the EP code with different settings and compilers.
}
\end{figure}
%%
%%%%%%%%%%%%%%%%%%%%%%%

The results shown in Fig. \ref{fig6} are obtained using the $\tt dp $ and $\tt tole = 0$. However, the use of different real number types and compilers can also influence the calculation time. Specifically, Fig. \ref{fig8} shows the CPU time obtained from the EP calculations conducted with the $\tt qp $ and $\tt dp$ using the Gfortran (Fig. \ref{fig8}(a) and (b)), and both Gfortran and Ifort compilers (Fig. \ref{fig8}(c)). 
Different options of the tolerance and pairing strength such as $\tt tole=0$ and  $10^{-8} $, and 
   \textbf{G} = matrix ($G_{jj'}$) or scalar ($G$) 
are applied. When using the lower precision $\tt dp$, the calculation time is reduced by approximately 80 times compared to using the $\tt qp$ (see Fig. \ref{fig8}(a) and (c)). On the other hand, using $\tt tole=10^{-8}$ is on average 1.5 times faster than $\tt tole=0$ (see Fig. \ref{fig8}(b)). Moreover, employing the Ifort compiler reduces the calculation time by about 1.3 times compared to using Gfortran with the $\tt dp$. However, when using the highest precision as $\tt dp$, the performance of two compilers are nearly the same (see Fig. \ref{fig8}(c)). The use of different types of \textbf{G} (matrix/scalar) has almost no effect on the running time (see Fig. \ref{fig8}(a) and (b)). To use the solutions obtained from the EP code as input for nuclear structure calculations such as mean field approach or shell model, users can opt for the configuration with the lowest precision to save the computation time while still ensuring the quality of the results, i.e., the standard settings are $\tt dp$ and $\tt tole=10^{-8}$.

\section{Conclusion and outlook}
\label{Con}

The EP code (v1.0) provides exact solutions to the nuclear pairing problem at zero temperature, accommodating configurations with either even (seniority-zero space) or odd (seniority-one space) numbers of nucleons. The sparsity and symmetry of the pairing matrix are utilized to optimize the computational time. Construction of the sparse pairing matrix is performed quickly and efficiently 
by representing the occupied states of nucleons in the binary representation. 
Binary algorithms within subroutines 
\textbf{gens}, 
\textbf{npairing}, 
\textbf{genp}, 
\textbf{hashs},  
\textbf{gind} 
swiftly generate, evaluate, extract, locate, and store the non-zero elements in the upper half of the pairing matrix, 
significantly reducing the matrix construction time.

Diagonalization of the pairing matrix is flexibly and efficiently handled by the ARPACK package for large dimensions and 
the LAPACK package for smaller dimensions. 
A technique to exactly identify the block containing the ground state in odd configurations has been introduced, 
substantially diminishing the calculation time for configurations with an odd number of nucleons. Furthermore, the Kahan compensation algorithm is applied to address the compensated summation, achieving the higher error stability in calculations.

The EP code can serve as 
a versatile alternative to the conventional BCS calculations used within the mean-field approximation or shell model. 
The current code is easy to use by specifying a set of input, 
i.e., parameters such as the number of single-particle levels ($ \Omega $), 
the corresponding single-particle energies ($ \epsilon_j $), 
the number of nucleons ($ N $), 
and the pairing strength ($ \textbf{G} = \text{matrix/scalar} $), 
enabling flexible input changes by users. In case \textbf{G} is a matrix, the user can directly enter either the elements $G_{jj'}$ or the pairing matrix elements $V(jj, j'j')$. With $ \Omega $ less than 26, 
the EP calculations typically complete in less than one second to a few hours. Moreover, the current code is designed to facilitate straightforward extensions to finite-temperature calculations in the future.

\section*{Declaration of competing interest}

The authors declare that they have no known competing financial interests or personal relationships that could have appeared to influence the work reported in this paper.

\section*{Data availability}

The EP source code is available at: \url{https://github.com/ifas-mathphys/epcode_v01}. Regular updates and bug ﬁxes will be provided.

\section*{Acknowledgements}

The authors (T.Q.V,  L.T.P, T.V.D, and N.Q.H) express their gratitude for all the valuable support from Duy Tan University (DTU).



%% References


%\bibliographystyle{elsarticle-num}



\end{document}